\input harvmac.tex
\let\includefigures=\iftrue
\newfam\black
\includefigures
\input epsf
\def\figin{\epsfcheck\figin}\def\figins{\epsfcheck\figins}
\def\epsfcheck{\ifx\epsfbox\UnDeFiNeD
\message{(NO epsf.tex, FIGURES WILL BE IGNORED)}
\gdef\figin##1{\vskip2in}\gdef\figins##1{\hskip.5in}
\else\message{(FIGURES WILL BE INCLUDED)}%
\gdef\figin##1{##1}\gdef\figins##1{##1}\fi}
\def\DefWarn#1{}
\def\figinsert{\goodbreak\midinsert}
\def\ifig#1#2#3{\DefWarn#1\xdef#1{fig.~\the\figno}
\writedef{#1\leftbracket fig.\noexpand~\the\figno}%
\figinsert\figin{\centerline{#3}}\medskip
\centerline{\vbox{\baselineskip12pt
\advance\hsize by -1truein\noindent
\footnotefont{\bf Fig.~\the\figno:} #2}}
\bigskip\endinsert\global\advance\figno by1}
\else
\def\ifig#1#2#3{\xdef#1{fig.~\the\figno}
\writedef{#1\leftbracket fig.\noexpand~\the\figno}%
\centerline{\vbox{\baselineskip12pt
\footnotefont{\bf Fig.~\the\figno:} #2}}
\global\advance\figno by1}
\fi
\def\co{{\cal O}}

\def\cn{{\cal N}}

\def\pr{{\prime}}
\def\cl{{\cal L}}
\def\cj{{\cal J}}
\def\ct{{\cal T}}

\def\p{{\bf{P}}}

\def\cyt{Calabi--Yau threefold}
\def\cym{Calabi--Yau manifold}
\def\cyf{Calabi--Yau fourfold}
\def\ie{{\it i.e.\/},}
\def\eg{{\it e.g.\/},}
\Title{\vbox{\baselineskip12pt\hbox{hep-th/9903104}
\hbox{IASSNS-HEP-99-28}}}
{\vbox{
\centerline{Geometrical Aspects of Fivebranes in}
\vskip 10pt
\centerline{Heterotic/F-Theory Duality in Four Dimensions}}}
\vskip 10pt
\centerline{Duiliu-Emanuel Diaconescu and Govindan Rajesh}
\medskip
\centerline{\it School of Natural Sciences,}
\centerline{\it Institute for Advanced Study,}
\centerline{\it Olden Lane,}
\centerline{\it Princeton, NJ 08540}
\centerline{\tt diacones, rajesh@ias.edu}
\medskip
\bigskip
\centerline{\bf Abstract}
\noindent
We use the method of stable degenerations to study the local geometry of
\cyf s for F-theory compactifications dual to heterotic compactifications
on a \cyt\ with fivebranes wrapping holomorphic curves in the threefold.
When fivebranes wrap intersecting curves, or when many fivebranes wrap the
same curve, the dual fourfolds degenerate in interesting ways. We find 
that
some of these can be usefully described in terms of degenerations of the 
base of the elliptic fibrations of these fourfolds. We use Witten's 
criterion
to determine which of the fivebranes can lead to the generation of a
non-perturbative superpotential.
\Date{March, 1999}

\newsec{Introduction}
Heterotic compactifications on elliptic \cyt s provide us with
phenomenologically interesting vacua in four dimensions with $N=1$
supersymmetry. Moreover, such vacua are conjectured to have a dual 
description
in terms of F-Theory compactified on elliptically fibered \cyf s.
The heterotic vacua are specified by a choice of gauge bundle, 
constructed using the general techniques of Friedman, Morgan and Witten
\ref\FMW{R. Friedman, J. Morgan, E. Witten, ``Vector Bundles And F
Theory'', Commun. Math. Phys. {\bf 187} (1997) 679, hep-th/9701162.}. 
It was
shown there that such vacua often include fivebranes wrapping the
elliptic fibers of the \cyt\ $Z$, whose number is determined by an 
anomaly 
cancelation condition. 

More generally, we can have fivebranes on the heterotic side wrapping
holomorphic curves in $Z$. In this paper, we examine the geometry of 
fourfold
duals of heterotic vacua with fivebranes. 
The cohomology classes of the curves being wrapped are fixed by the 
general
heterotic anomaly cancelation condition 
\eqn\anomaly{[W]= c_2(TZ) - \lambda(V_1) - \lambda(V_2),}
where $[W]$ is the class of the wrapped curves, $c_2(TZ)$ is the second 
Chern class of the tangent bundle of the \cyt\ $Z$, and $\lambda(V_i)$ 
are the second Chern classes of the vector bundles on $Z$. Note that 
this only fixes the curves up to their cohomology classes. 
Given that $Z$ is elliptically fibered with a section $\sigma$, we see 
that the
class $[W]$ may be written $[W] = C_1\sigma + C_2$, so that under the
projection $\pi : Z \rightarrow B_2$, $C_1$ maps to a divisor in $B_2$, 
and $C_2$
maps to $h[p]$, where $[p]$ is the class of a point in $B_2$ and
$h=\int_{\sigma}{C_2}$ is an integer. Thus the class $C_2$ actually 
describes
the fivebranes wrapping the elliptic fiber of $Z$, while $C_1\sigma$ 
describes
the fivebranes wrapping holomorphic curves in the base $B_2$. In this 
paper,
we will refer to the first kind (\ie\ those in the class $C_2$) as vertical
fivebranes, and the second (\ie\ those in the class $C_1\sigma$) as
horizontal ones. (It is of course possible for a fivebrane to wrap a curve
which has both vertical 
(\ie\ fiber) components and horizontal (\ie\ base) components.)

Under the duality, the vertical fivebranes map to F-theory threebranes
\ref\AC{B. Andreas and G. Curio, ``Three-Branes and Five-Branes in 
N=1 Dual String Pairs'', Phys. Lett. {\bf B417} (1998) 41,
hep-th/9706093.}. The number of the F-theory threebranes is related to the
Euler number of the fourfold by tadpole anomaly cancelation
\ref\SVW{S. Sethi, C. Vafa and E. Witten, ``Constraints on 
Low-Dimensional String Compactifications'', Nucl. Phys. {\bf B480} (1996) 
213, hep-th/9606122.}.
This relation is actually modified in the presence of the
four form field strength $G$. Also, when the threebranes coincide with the
sevenbranes wrapping divisors in the base $B_3$ of the fourfold over 
which
the elliptic fibration degenerates, they behave as instantons, breaking 
the gauge group observed from the singularity to a smaller one. For the 
purposes
of this paper, we will ignore both these possibilities, since they play no
role in our analysis. Our results will be valid even in the presence of 
these complications. 

The horizontal fivebranes, on the other hand, map to geometric data on the
F-theory side. Specifically, if a fivebrane wraps a curve $C$ 
in $B_2$, then the F-theory base $B_3$ is blown up once over the 
corresponding curve
\nref\GR{G. Rajesh, ``Toric geometry and F-theory/Heterotic Duality in 
Four Dimensions'', JHEP {\bf 12} (1998) 018, hep-th/9811240.}%
\nref\BM{P. Berglund and P. Mayr, ``Heterotic String/F-theory 
Duality from Mirror Symmetry'', hep-th/9811217.}%
according to \GR\ and independently \BM.
Fivebranes wrapping the same curve correspond to an equal number of 
blow-ups on the F-theory side.

For the purposes of this paper, we choose to ignore the 
vector bundles on the heterotic theory altogether, and concentrate 
instead on the local physics of the fivebranes. Thus, we consider an 
extreme
situation when the bundles $V_{1,2}$ are without structure group. 
This is analogous to the six dimensional vacuum with $24$ small 
instantons, the anomaly cancelation being entirely due to fivebranes.
The base of the F-Theory threefold in this case acquires several blowups
\ref\MV{D.R. Morrison and C. Vafa, ``Compactifications of 
F-Theory on Calabi--Yau Threefolds -- I'', Nucl.Phys. {\bf B473}
(1996) 74, hep-th/9602114; ``Compactifications of F-Theory on 
Calabi--Yau Threefolds -- II'', Nucl.Phys. B476 (1996) 437,
hep-th/9603161.}, whose local
description involves the method of stable degenerations
\nref\tasi{P.S. Aspinwall, ``K3 Surfaces in String Duality'', TASI(96), 
hep-th/9611137.}%
\nref\Korb{P.S. Aspinwall and D. R. Morrison, ``Point-like Instantons 
on K3 Orbifolds'', Nucl. Phys. {\bf B503} (1997) 533, hep-th/9705104.}%
\refs{\tasi, \Korb}.

In the four dimensional situation, it is precisely the horizontal 
fivebranes
that correspond to blowup modes in the fourfold base. In this paper,
will use the method of stable degenerations to describe this geometry.
After a brief description of the general technique in section 2, we 
explicitly
work out the fourfold geometry in the stable degeneration limit for
a single horizontal fivebrane in section 3.    
When two fivebranes wrap intersecting curves, or when several
fivebranes
wrap the same curve, the dual fourfolds degenerate in interesting ways.
We find that it is more useful to describe these degenerations by studying
the degenerations of the corresponding base $B_3$, which 
is the subject of section 4. We find, for example, that when two horizontal
fivebranes intersect, the base $B_3$ acquires a conifold singularity, while
$k$ horizontal fivebranes wrapping the same curve $C$ in $B_2$ lead to an
$A_{k-1}$ singularity fibered over the corresponding curve in $B_3$.

In addition to affecting the geometry of the
fourfolds, the fivebranes can contribute to the nonperturbative
superpotential. Section 5 is devoted to discussing the criteria for
determining which fivebranes can contribute to the superpotential, based on the
work of
\nref\W{E. Witten, ``Non-Perturbative Superpotentials in String Theory'',
Nucl. Phys. {\bf474} (1996), hep-th/9604030.}%
\nref\G{A. Grassi, ``Divisors on Elliptic Calabi-Yau 4-Folds and The 
Superpotential in F-Theory, I'', alg-geom/9704008.}%
\refs{\W, \G}.

\newsec{Heterotic Models and Stable Degenerations}

This section consists of a brief review of heterotic vacua and 
the stable degeneration limit. 
Consider a general $d=4$, $N=1$ heterotic vacuum specified by 
the compactification data $(Z,V_1,V_2)$. Here 
$\pi:Z\rightarrow B_2$ is a nonsingular elliptic Calabi-Yau threefold
with a section $\sigma:B_2\rightarrow Z$. 
$V_1,V_2$ are two holomorphic 
bundles with structure group $G_1^c,G_2^c$ where $G_{1,2}\subset E_8$.
Throughout the paper, the base $B_2$ will be taken to be 
a Hirzebruch surface $F_e$, with $e=0,1,2$ in order to insure the 
smoothness of the total space Z. Moreover, the present considerations are
restricted to heterotic models which admit an F-theory dual. Therefore,
$Z$ will be taken to be a smooth Weierstrass model 
\eqn\WA{
zy^2=x^3-axz^2-bz^3}
in $\p\left(\co_B\oplus\cl^2\oplus\cl^3\right)$ with $\cl\simeq
K_{B_2}^{-1}$ in order to satisfy the Calabi-Yau condition. $a,b$ are
sections of $\cl^4,\cl^6$.
The bundles $V_{1,2}$ will be specified by spectral data 
$\left(\Sigma_{1,2},\cn_{1,2}\right)$ 
\nref\BJPS{M. Bershadsky, A. Johansen, T. Pantev, V. Sadov, 
``On Four-Dimensional Compactifications of F-Theory'', 
Nucl. Phys. {\bf B505} (1997) 165, hep-th/9701165.}%
\nref\FMWb{R. Friedman, J.W. Morgan, E. Witten, ``Vector Bundles 
over Elliptic Fibrations'', alg-geom/9709029.}%
\refs{\FMW,\BJPS,\FMWb}.

According to \refs{\MV, \FMW}, the dual F-theory model can be 
constructed by taking the size of the base $B_2$ very large 
so that we can use adiabatic arguments. The elliptic fibration 
$\pi:Z\rightarrow B_2$ is then replaced adiabatically 
by a $K3$ fibration over $B_2$ with total space a Calabi-Yau 
fourfold $X$. It turns out that $X$ can be represented as an 
elliptic Weierstrass model $\pi^\pr:X\rightarrow B_3$ where 
$p:B_3\rightarrow B_2$ is a rationally ruled threefold over $B_2$.
The moduli map between the heterotic and F-theory data 
has been discussed intensively in 
\nref\CD{G. Curio and R.Y. Donagi, ``Moduli in N=1 heterotic/F-theory 
duality'', Nucl. Phys. {\bf B518} (1998) 603, hep-th/9801057.}%
\refs{\MV, \FMW, \BJPS, \CD, \GR, \BM}.

A particularly convenient approach is the method 
of stable degenerations 
\nref\PSA{P.S. Aspinwall, ``Aspects of the Hypermultiplet Moduli Space 
in String Duality'', J. High Energy Phys. {\bf 04} (1998) 019,
hep-th/9802194.}%
\refs{\FMW, \Korb, \PSA} which establishes a direct geometric 
correspondence between the two sets of data. Briefly, this consists 
of taking a limit in which the size of the elliptic fiber of $Z$ is also 
very large. Then, the F-theory fourfold $X$ 
degenerates to a union of two fourfolds $X=X_1\sqcup_Z X_2$ 
glued together along a three dimensional variety isomorphic to 
$Z$. Both $X_1,X_2$ are elliptically fibered over rationally ruled 
threefolds isomorphic to $B_3$
with projections $\pi^\prime_1, \pi^\prime_2$ such that 
$\pi=\pi^\prime_1|_Z=\pi^\prime_2|_Z$. The composite 
maps $p\circ \pi^\prime_{1,2}$ are fibrations of $X_{1,2}$ over $B_2$, 
with generic fiber a rational elliptic surface $dP_9$.

As explained in {\Korb, \PSA} this procedure is very useful 
for studying the F-theory local geometry associated to 
heterotic small instantons. However, since the stable degeneration 
is, strictly speaking, at infinite distance in the moduli space metric,
it may not be suitable for describing particular 
physical processes. This will be the case with certain fivebrane 
interactions and nonperturbative instanton effects. These phenomena 
are better described using a smooth resolution of the Calabi-Yau 
fourfold in a region of the moduli space where the geometric map 
developed via stable degenerations is still valid.

\newsec{A Single Horizontal Fivebrane}

Here we consider the case of a single heterotic fivebrane wrapping a
nonsingular irreducible curve $C\subset B_2$. The goal of this 
section is to determine the corresponding F-theory geometry in the 
stable degeneration limit. Recall that the fourfold $X\equiv X_1$ 
can be represented as a Weierstrass model over a base $B_3$ which 
can be identified with the total space of a projective bundle 
$\p\left(\co_{B_2}\oplus\ct\right)$
over $B$. 
In the following, we will restrict
to models in which $\ct \simeq \co_{B_2}\left(-\Gamma\right)$ 
for some effective divisor $\Gamma$ on $B_2$. To introduce some notation,
note that the fibration $p:B_3\rightarrow B_2$ has two disjoint 
holomorphic sections which will be denoted by $S_0,S_\infty$ by analogy 
with Hirzebruch surfaces. Then
\eqn\ruling{\eqalign{
& S_\infty=S_0+p^*\Gamma\cr
& K_{B_3}=-2S_0+p^*\left(K_{B_2}-\Gamma\right).\cr}}
The fourfold 
$\pi^\pr : X\rightarrow B_3$ is described as a Weierstrass model
\eqn\WB{
y^2=x^3-fx-g}
in $\p\left(\co_{B_3}\oplus{\cl^\pr}^2\oplus{\cl^\pr}^3\right)$, 
with $\cl^\pr=\co_{B_3}\left(S_0\right)\otimes p^*K_{B_2}^{-1}$. 
Note that the 
total space $X$ is not a Calabi-Yau variety in this case. 
As in 
\refs{\Korb,\tasi}, the structure of the elliptic fibrations is 
determined by the divisors
\eqn\divisors{\eqalign{
& F=4S_0-4p^*K_{B_2}\cr
& G=6S_0-6p^*K_{B_2}\cr
&\Delta =12S_0-12p^*K_{B_2}.\cr}}
The base $B_3$ is glued to the base of the second fourfold of the 
stable degeneration along the section $S_\infty$. Therefore the gluing 
divisor is the restriction of the elliptic fibration to $S_\infty$
which can be described as a Weierstrass model with line bundle 
$K_{B_2}^{-1}$.
This generically defines a smooth Calabi-Yau space which is identified 
with $Z$. Also, note that the restriction of the elliptic fibration 
to the generic rational fiber of $p:B_3\rightarrow B_2$ has exactly 
$12$ $I_1$ fibers, therefore it is a rational elliptic surface as 
claimed before. 

Since we are interested in heterotic models without structure group, 
we first enforce a section of $II^*$ singularities along the section 
$S_0$ of $B_3$ ensuring an unbroken $E_8$ group. This corresponds to 
splitting the divisors $F,G,\Delta$ into
\eqn\splitting{\eqalign{
& F=4S_0+F^\pr,\qquad F^\pr=-4p^*K_{B_2}\cr
& G=5S_0+G^\pr,\qquad G^\pr=S_0-6p^*K_{B_2}\cr
& \Delta = 10S_0+\Delta^\pr,\qquad \Delta^\pr=2S_0-12p^*K_{B_2}.\cr}}
The location of the small instantons will be determined by the collision
of the component $\Delta^\pr$ of the discriminant with the section $S_0$. 
A more precise description requires the explicit expression of the 
polynomials $f,g,\delta$ function of the coordinates of the base. 
In fact, the Weierstrass model can be characterized in a neighborhood 
of $S_0\simeq B_2$ as a hypersurface 
\eqn\hyper{
y^2=x^3-fx-g}
in the bundle 
$\ct\oplus\ct^2\otimes\cl^2\oplus\ct^3\oplus\cl^3$ \FMW.
Let $s$ denote an affine coordinate along the fibers of $B_3$ in the 
neighborhood of $S_0$. Hence $s$ is a section of the normal bundle
$N_{S_0/B_3}\simeq \ct$. Then we have the 
following power series expansions
\eqn\series{\eqalign{
&f=s^4f_4\equiv s^4 f^\pr\cr
&g=s^5\left(g_5+sg_6\right)\equiv s^5 g^\pr\cr
&\delta=s^{10}\left[4s^2f_4^3+27
\left(g_5+sg_6\right)^2\right]\equiv s^{10} \delta^\pr\cr}}
where $f_4$ is a section of $\cl^4$ and 
$g_5,g_6$ are sections of $\cl^6\otimes\ct,\cl^6$. 

The component $\Delta^\pr$ is defined by $\delta^\pr =0$. This is a 
quadratic equation in $s$ with discriminant (up to a multiplicative 
factor)
\eqn\quadr{
f_4^3g_5^2.}
Therefore $\Delta^\pr$ is a double cover of $B_2$ branched along the 
locus 
\eqn\branch{
f_4=g_5=0.}
For future reference, let $A,C$ denote the loci $f_4=0$ and 
$g_5=0$ in $S_0\simeq B_2$ respectively. By a suitable choice of the 
line bundle $\ct$, these can be assumed nonsingular irreducible 
curves. 
Note that $C$ actually represents the locus of intersection of 
$\Delta^\pr$ with $S_0$, which is precisely the location of 
the small instanton. The singularity of the total space $X$ 
induced by the $I_1+II^*$ collision along $C$ requires 
a blow-up of the base, as will be shortly detailed. The points 
of the discriminant in the inverse image of $A$ in $\Delta^\pr$
represent a locus of $II$ elliptic fibers of $X$. 
Furthermore, note that $\Delta^\pr$ intersects the section $S_\infty$ 
along the locus $L$ given by 
\eqn\inters{
4f_4^3+27g_6^2=0}
which is a section of $-12K_{B_2}$. The equation $\delta^\pr=0$ is 
identically satisfied when 
\eqn\simult{
4f_4^3+27g_6^2=0,\qquad g_5=0}
hold simultaneously. Therefore, the discriminant $\delta^\prime$
contains the $-12K_{B_2}\cdot C$ fibers of $p:B_3\rightarrow B_2$ 
localized at the points of intersection $C\cdot L$ in $B_2$.

As observed above, a smooth model of the fourfold $X$ can be constructed 
by blowing-up the base $B_3$ along the curve $C$ embedded in $S_0$.
In order to describe the resulting configuration, let $(t,u)$ be 
local coordinates on $S_0$ near a point $P$ of $C$ so that $t$ is
a normal coordinate and $u$ is a coordinate on $C$ centered at $P$. 
The section $g_5$ is assumed to have a simple zero along $C$, 
therefore, it will have a local expansion of the form 
\eqn\locexp{
g_5=tg_5^\pr,}
where $g_5^\pr$ does not vanish at $t=0$. 
The blow-up is described in affine coordinates by 
setting $s=s_1t_1, t=t_1$ which results in 
\eqn\blowup{\eqalign{
&f=s_1^4t_1^4f_4\cr
&g=s_1^5t_1^6(g^\pr_5+s_1g_6)\cr
&\delta=s_1^{10}t_1^{12}\left[4s_1^2f_4^3+27\left(
g_5^\pr+s_1g_6\right)^2\right].\cr}}
The Weierstrass model can be set in normal form by rescaling the 
coordinates $x,y$ by appropriate powers of $t_1$ obtaining
\eqn\rescaling{\eqalign{
&f=s_1^4f_4\cr
&g=s_1^5(g_5+s_1t_1g_6)\cr
&\delta=s_1^{10}\left[4s_1^2f_4^3+27\left(
g_5^\pr+s_1g_6\right)^2\right].\cr}}
Globally, the exceptional divisor $D$
is isomorphic to the projective bundle 
$\p\left(N_{C /{B_3}}\right)$ whose $P^1$ fiber is parameterized 
by the affine coordinate $s_1$. The degree of the ruling is 
$-6K_{B_2}\cdot C$. 
The strict transform of the vertical divisor $p^*C$ is isomorphic to 
another 
rationally ruled surface $D^\pr$ over $C$. Let $C_0^D, C_\infty^D$, 
$C_0^{D^\pr}, C_\infty^{D^\pr}$ denote the disjoint sections of $D,D^\pr$ 
respectively and let $S_0^\pr$ denote the strict transform of the 
section $C_0$. Then $S_0^\pr$ and $D$ intersect along $C_0^D$, 
and $D$ and $D^\pr$ intersect along a common section $C_0^D\simeq 
C_\infty^{D^\pr}$. The blown-up base ${\bar B}_3$ can be regarded as a 
fibration ${\bar p}:{\bar B}_3\rightarrow B_2$ with generic fiber 
$P^1$. Above $C\subset B_2$, the fiber consists of two rational 
components intersecting transversely. See Fig.1. for a schematic 
representation.

\ifig\figA{The blow-up geometry corresponding to a single horizontal 
fivebrane wrapped around the curve $C$. $D$ and $D^\pr$ are the 
exceptional divisor and the proper transform of $p^*C$ 
respectively. The thick lines represent the intersection of the 
discriminant with $D$, $D^\pr$.}{\epsfxsize3.0in\epsfbox{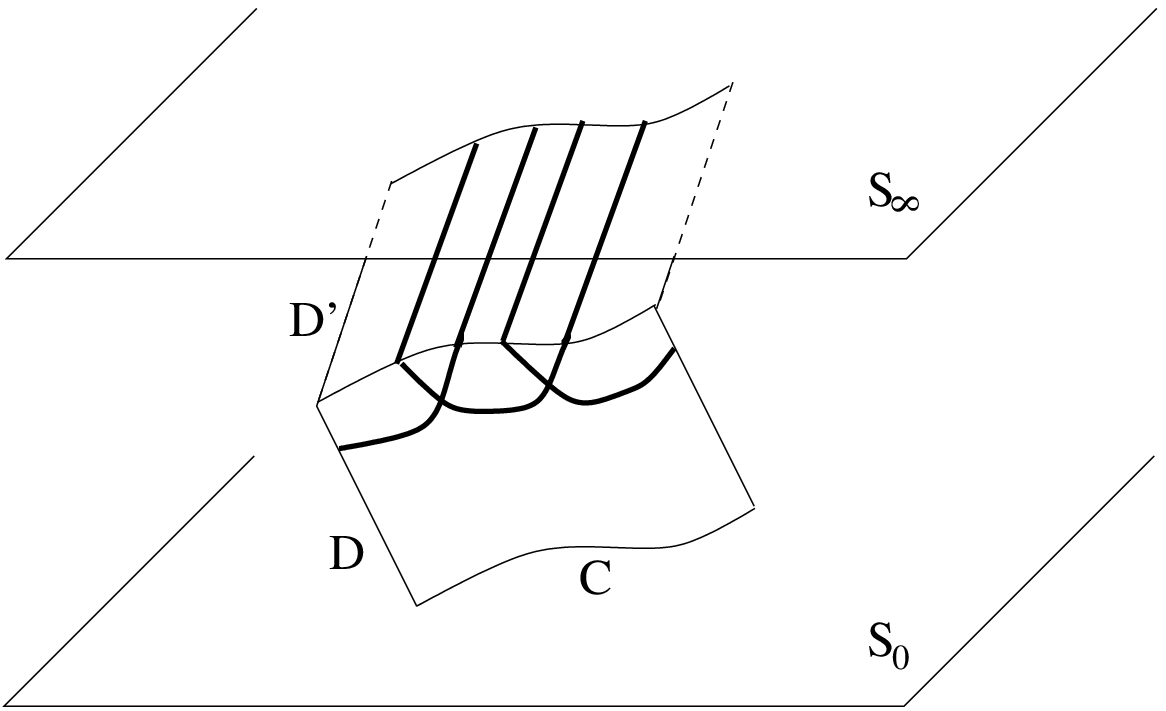}}

It follows from \rescaling\ that the proper transform 
of the discriminant ${\bar\Delta}^\pr$ intersects $D$ 
along a divisor in the 
class $2C_\infty^D$ which does not meet $S_0^\pr$. Therefore 
$S_0^\pr$ is an isolated section of $II^*$ fibers of the resulting 
fourfold $\bar X$. Furthermore, ${\bar\Delta}^\pr$ intersects 
the surface $D^\pr$ along $-12K_{B_2}\cdot C$ $P^1$ fibers which carry 
$I_1$ elliptic fibers. An interesting exception arises in the case 
$B_2\simeq F_2$ and $C=C_0$. In this case, the exceptional divisor 
$D$ is simply $P^1\times P^1$ since $K_{B_2}\cdot C_0=0$ and 
${\bar\Delta}^\pr$ intersects $D$ along two disjoint sections. 
The strict transform $D^\pr$ does not intersect ${\bar\Delta}^\pr$ 
and there are no vertical lines carrying $I_1$ elliptic fibers.

If everything else is generic, the elliptic 
fibration is smooth away from $S_0^\pr$ and it has a locus of type $II$ 
fibers which projects to the curve $A$ in the base. From here, a 
smooth model is easily obtained by blowing up $\bar X$ along the section 
$S_0^\pr$, obtaining an $E_8$ Hirzebruch-Jung tree fibered over 
$S_0^\pr$. 
These exceptional divisors play an important role in the dynamics 
of $d=3$ $N=2$ pure gauge theories 
\nref\KV{S. Katz, C. Vafa, ``Geometric Engineering of N=1 Quantum Field 
Theories'', Nucl. Phys. {\bf B497} (1997) 196, hep-th/9611090.}%
\nref\CV{C. Vafa, ``On N=1 Yang-Mills in Four Dimensions'', 
Adv. Theor. Math. Phys. {\bf 2} (1998) 497, hep-th/9801139.}%
\refs{\KV, \CV}.

\subsec{The Spectral Cover}

The heterotic spectral cover corresponding to the above fourfold 
degeneration can be determined by analogy with the six dimensional 
situation considered in \refs{\Korb,\PSA}. In that case, it is known 
\MV,\ that the extra K\"ahler moduli associated to a blow-up in the 
base corresponds to an extra $(1,0)$ tensor multiplet. This allows 
an identification of the threefold degeneration with a small $E_8$ 
instanton, that is a heterotic fivebrane. In the present situation, 
the low energy effective action corresponding to the fourfold 
degeneration can be easily derived regarding the model as a IIB 
compactification on the base ${\bar B}_3$ with a varying 
dilaton. Note that blowing-up $B_3$ along the curve $C$ 
has the effect of producing nontrivial homology 3-cycles. More precisely, 
the intermediate Jacobian $J\left({\bar B}_3\right)$ is isomorphic 
to the Jacobian $J\left(C\right)$ 
\nref\CG{C.H. Clemens and P.A. Griffiths, ``The Intermediate Jacobian 
of The Cubic Threefold'', Ann. of Math. (2) {\bf 95} (1972) 281.}%
\nref\Ty{A.N. Tyurin, ``Five Lectures on Three-Dimensional Varieties'',
Russian. Math. Surveys. (5) {\bf 27} (1972) 1.}%
\refs{\CG,\Ty}.
Therefore there is a $1-1$ correspondence between 
$H^{1,2}\left({\bar B}_3\right)$ and $H^{0,1}\left(C\right)$
given by the cylinder (or equivalently, Abel-Jacobi) map of 
\refs{\CG,\Ty}. Note that $h^{0,3}\left({\bar B}_3\right)=
h^{0,3}(B_3)
=0$, therefore $J\left({\bar B}_3\right)$ is in this case a 
principally polarized abelian variety. 

The low energy effective action is then determined by reducing the 
$SL(2,Z)$ invariant 4-form $C^{(4)}$ with self-dual field strength 
$G^{(5)}$ along the elements of $H^{1,2}\left({\bar B}_3\right)$.
The discussion is similar to 
the reduction of the M-theory fivebrane along a compact Riemann surface 
found in 
\ref\W{E. Witten, ``Solutions of Four-Dimensional Field Theory via 
M Theory'', Nucl. Phys. {\bf } (1997) , hep-th/9703166.}.
We have an ansatz 
\eqn\anszatz{
G^{(5)}=F\wedge \Lambda + \ast F\wedge\ast\Lambda}
where $F$ is a 2-form on $R^4$ and $\Lambda$ is a 3-form on 
${\bar B}_3$.
The equation of motion $dG^{(5)}=0$ yields the Maxwell equations for $F$
and requires $\Lambda$ to be a harmonic 3-form. It follows from Hodge 
theorem 
that $\Lambda$ defines a point in the intermediate Jacobian 
$J\left({\bar B}_3\right)$. The corresponding low-energy 
effective action 
consists of $U(1)^g$ massless gauge fields whose couplings are 
determined by 
$J\left({\bar B}_3\right)$ ($g$ is the genus of the wrapped Riemann
surface). Taking into account the identification 
$J\left({\bar B}_3\right)\simeq J\left(C\right)$, we conclude that 
the effective action derived this way is in fact identical with that of 
an M-theory fivebrane wrapping the compact Riemann surface $C$
\nref\LOW{A. Lukas, B.A. Ovrut, D. Waldram, ``Non-standard embedding 
and five-branes in heterotic M-Theory'', hep-th/9808101.}%
\nref\Don{R. Donagi, A. Lukas, B.A. Ovrut, D. Waldram, 
``Non-Perturbative Vacua and Particle Physics in M-Theory'', 
hep-th/9811168.}%
\refs{\LOW,\Don}.
This provides 
physical evidence for identifying the fourfold degeneration with an $E_8$ 
heterotic fivebrane wrapping the curve $C$ in the base $B_2$. 

A more precise geometric picture can be achieved as follows. Note that 
the 
strict transform $D^\pr$ discussed in the previous subsection intersects 
the section $S_\infty$ along a curve isomorphic to $C$. The 
restriction
of the elliptic fibration to $D^\pr$ gives an elliptic threefold 
$Q\rightarrow D^\pr$ with $I_1$ degenerations along $-12K_{B_2}\cdot C$ 
rational fibers of $D^\pr$. Moreover, the heterotic Calabi-Yau threefold 
$Z$ is identified with the restriction of 
${\bar \pi}^\pr : {\bar X}\rightarrow {\bar B}_3$
to $S_\infty$. Therefore $Q$ and $Z$ meet along a surface $S$ 
elliptically fibered over $C$ and with $-12K_{B_2}\cdot C$ 
$I_1$ fibers. As in \PSA,\ this is an irreducible component of the 
spectral cover $\Sigma=\sigma\cup S$. Note that the reducible spectral 
cover describes a ``bundle without structure group'' which is 
in proper language the ideal sheaf $\cj_{C/Z}$. Similar degenerations 
have been also considered in 
\ref\ES{E. Sharpe, ``Extremal Transitions in Heterotic String Theory'',
Nucl. Phys. {\bf B513}, 175, hep-th/9705210.}.

Before further developing the subject by considering multiple small 
instantons, a couple of remarks are in order. Some aspects of the 
present construction may be better understood by comparison with 
the threefold degenerations studied in \PSA\ (section 3.4). Note that 
in that case, the role of the elliptic threefold $Q$ is played by an 
elliptic surface which is generically trivial i.e. 
$Q\simeq P^1\times T^2$. 
The injective map $H^1\left(Q,Z\right)\rightarrow H^3\left(X,Z\right)$ 
induces nontrivial homology 3-cycles on $X$. The spectral cover is again 
reducible $\Sigma=\sigma\cup S$ where $S$ is isomorphic to the constant 
elliptic fiber of $Q$. Therefore, the Jacobian of $\Sigma$, which is 
isomorphic
to that of $S$, is mapped injectively into the intermediate Jacobian of 
$X$. This is part of the Heterotic/F-theory map which maps the 
position of the small instanton along the elliptic fiber to Ramond-Ramond
moduli of $X$. In the present situation, $S$ is an elliptic surface 
which can be written as a Weierstrass model over 
$C\subset\sigma$
with line bundle $K_{B_2}^{-1}|_C$. This shows that the normal bundle 
$N_{C/S}\simeq K_{B_2}|_{C}$ is of negative degree, therefore
$C$ cannot be deformed in $S$. Accordingly, there are no moduli 
parameterizing the position of the small instanton along the elliptic 
fiber. 
If the genus of $C$ is $g\geq 1$, the surface $S$ has a nontrivial 
Jacobian isomorphic to that of $C$ which injects in the Jacobian 
of $X$ as explained before. This is again part of the Heterotic/F-theory 
map having, however, a different physical interpretation in terms of the 
effective couplings of the four-dimensional low-energy action.

Note that there is an exception to this generic behavior. Namely, as 
also noted before, if $B\simeq F_2$ and $C=C_0$, it turns out that the 
threefold $Q$ and the surface $\Sigma=Q\cap Z$ are trivial elliptic 
fibrations, that is $Q\simeq D^\pr \times T^2$ and 
$\Sigma\simeq P^1\times T^2$. This behavior is very similar to the 
six dimensional case since the Jacobian of $\Sigma$ is now isomorphic 
to that of the trivial elliptic fiber. In particular, the fivebrane 
wrapped on $C\subset \sigma$ can be moved along the elliptic fiber 
and its position on $T^2$ is parameterized by a point in the Jacobian.

\subsec{Interaction with Vertical Fivebranes -- A Puzzle}

We know from anomaly cancelation considerations that vertical fivebranes 
should also be present. Since they are obviously mobile along the 
heterotic base, the vertical fivebranes can collide the horizontal brane 
by a suitable tuning of moduli. For simplicity, we consider 
a single vertical fivebrane approaching the horizontal one.
The effective theory on the non-compact directions of the vertical 
fivebrane is a
free $U(1)$ gauge theory with three neutral complex chiral 
multiplets. Two chiral multiplets parameterize the motion of the 
fivebrane along the base. The scalar component of the third, $\Phi=
\phi +ia$, incorporates the position $\phi$ along the interval 
$S^1/Z_2$ 
and the fivebrane axion $da=*_4dB$. 

In flat eleven dimensional space, 
\nref\HK{A. Hanany, I. R. Klebanov, ``On Tensionless Strings in 
$3+1$ Dimensions'', Nucl. Phys. {\bf B482} (1996) 105, hep-th/9606136.}%
\nref\KOY{S. Kachru, Y. Oz and Z. Yin, ``Matrix Description of 
Intersecting M5 Branes, hep-th/9803050.}%
such a collision is expected to result in extra degrees of freedom 
-- ``tensionless strings'' -- localized on the intersection and 
eventually in an interacting superconformal fixed point \refs{\HK,\KOY}. 
In our case, the fivebranes 
wrap Riemann surfaces embedded in a curved space, therefore there could 
be extra effects and it is not clear if the decoupling takes place. 

A possible approach to this problem is via duality 
with F-theory. As noted above, the horizontal fivebranes map to 
background threebranes filling the non-compact $3+1$ directions. 
The expectation value of the field $\Phi$ is related to the 
position of the threebrane on the $P^1$ fiber of $p:B_3\rightarrow B$. 
In particular if the vertical fivebrane is localized at a point $P\in C$ 
on the base $B_2$, $\Phi$ corresponds to the threebrane position along 
the $P^1$ fiber of the proper transform $D^\pr$. Therefore, the collision
takes place precisely when the threebrane hits the exceptional divisor 
$D$. However, this apparently leads to a puzzle since the total space 
of the threefold $B_3$ is smooth hence, we have no reasons to expect an 
interacting theory on the threebrane worldvolume when it collides with 
the intersection $D^\pr\cap D$. In particular, although the heterotic
picture suggests a solitonic string of vanishing tension localized 
on the intersection of the fivebranes, no such object can be found 
in the threebrane worldvolume theory. At the present stage, although 
we do not 
have a complete solution of this puzzle, we suggest that it could be 
understood along the following lines. The present picture is valid 
within the framework of adiabatic duality, {\it i.e.}, the size of the 
base $B_2$ is much larger than the size of the elliptic fiber. 
Note that this is not the case in the stable degeneration limit, 
therefore the analysis is performed using a smooth F-theory model
in a suitable region of the moduli space so that the local geometry 
described by the stable degeneration is still valid. 
Then, the intersecting fivebranes can be locally modeled 
as two intersecting fivebranes in M-theory on $R^{1,8}\times T^2$. 
The vertical fivebrane corresponds to a fivebrane whose relative 
transverse directions are wrapped on $T^2$, while the horizontal 
fivebrane is transverse to the torus. Using standard duality arguments, 
this configuration can be mapped to a D3-brane in Type IIB theory 
moving in a smooth Taub-NUT background. Therefore the theory 
of the brane has a smooth moduli space, similarly to the situation 
considered in \ref\IR{N. Seiberg, ``IR Dynamics on Branes and Space-Time 
Geometry'', Phys. Lett. {\bf B384} (1996) 81, hep-th/9606017.}.
In particular, there are no singularities and no extra light degrees 
of freedom at any point in the moduli space. This is in agreement 
with the F-theory picture developed above.

Another question is related to the possibility of deforming the 
intersecting fivebranes, obtaining a single brane wrapping a smooth 
irreducible curve in the Calabi-Yau threefold. This is an interesting 
question also treated in \Don. For concreteness, let us consider 
a fivebrane wrapping an effective cycle in the homology class $C+nf$ 
where $f$ is the class of the elliptic fiber. As shown in a particular 
example in \Don,\ the moduli space of this fivebrane has, in general, 
many disconnected components. More precisely, let $\Sigma=\pi^*C$
and $i:\Sigma \rightarrow Z$ denote the inclusion. 
There could several homologically inequivalent configurations of 
holomorphic curves on $\Sigma$ mapping to the homology class 
$C+nf \in H_2(Z)$ under the map
$i_*:H_2(\Sigma)\rightarrow H_2(Z)$. However, although these 
curves are homologically equivalent when embedded in $Z$, they may not be
algebraically equivalent, in which case the different types of fivebranes 
cannot be deformed one to another in a supersymmetric way.
This gives rise to several distinct components of the fivebrane 
moduli space. 

For the class $C+nf$, one can show that there are no algebraic
(\ie\ holomorphic) deformations that lie entirely within $\Sigma=\pi^*C$.
However, it may be possible to deform this class in the ambient
space $Z$. Such deformations, cannot, however, exist if $C$ has no
deformations in $B_2$ (\eg\ if $C\cdot C < 0$). This is because any
deformation of $C+nf$ into a smooth irreducible curve $C'$ in $Z$
must project down to a deformation of $C$ in $B_2$, but $C$ cannot move in
$B_2$, which implies that the deformation lies entirely
in $\Sigma$, and hence does not exist. Thus, if there is a curve $C'$ in
$\Sigma$ which is different from $C+nf$ in $H_2(\Sigma)$, but maps to the same
element of $H_2(Z)$ under $i_*$, then we cannot deform holomorphically from
$C+nf$ to $C'$, so that they sit in disconnected components of the fivebrane
moduli space. However, it is not clear if such deformations
exist if $C$ moves in $B_2$. 
Deformations entirely within $\Sigma$ can exist when the multiplicity of the
horizontal  fivebranes is greater than one, \ie\ the fivebrane class is
$kC+nf$, with $k>1$. This phenomenon will be considered at a later stage. 

\newsec{Multiple Fivebranes}

\subsec{Intersecting Horizontal Fivebranes}

The discussion in the previous section can be generalized in several
different ways. We consider here the situation when we have two 
horizontal fivebranes wrapping two irreducible smooth curves
$C_1, C_2\subset B_2$ 
intersecting in a finite number of points. 
These curves may belong to
the same, or different cohomology class in the base. In \refs{\GR}, the
dual F-theory picture was described in terms of intersecting del Pezzo
surfaces. Here, we will describe it in terms of the geometry of the 
fourfold base. This corresponds to viewing F-theory as a 
nonperturbative Type IIB vacuum, which,
as we will see shortly, is a more useful picture.

For simplicity, we first consider the case when both $C_1, C_2$ are 
rational and they intersect in exactly one point (for example a fiber 
and a section of a Hirzebruch surface). 
The base of the fourfold ${\bar B}_3$ is now given by the  
$P^1$ bundle $p:B_3\rightarrow B$
blown-up twice along the curves $C_1$, $C_2$ embedded in the 
section $S_0$. Since the curves are intersecting the order of the 
blow-ups is important, as explained in the following. Assume that 
$C_1$ is blown-up first. 
Let $D_1,D_2$ denote the exceptional divisors, which are $P^1$ rulings 
over $C_1, C_2$ and let $E_1, E_2$ denote the classes of the 
fibers. Also, let $D_1^\pr, D_2^\pr$ denote the strict transforms of the 
vertical divisors $p^*C_1, p^*C_2$. Near the intersection 
point $P\in S_0$, the geometry looks as in Fig.2. 

\ifig\figB{The blow-up geometry corresponding to two intersecting 
horizontal fivebranes. Note that the fiber acquires a third component 
$E_3=D_1\cap D_2^\pr$ over the intersection point.}
{\epsfxsize3.0in\epsfbox{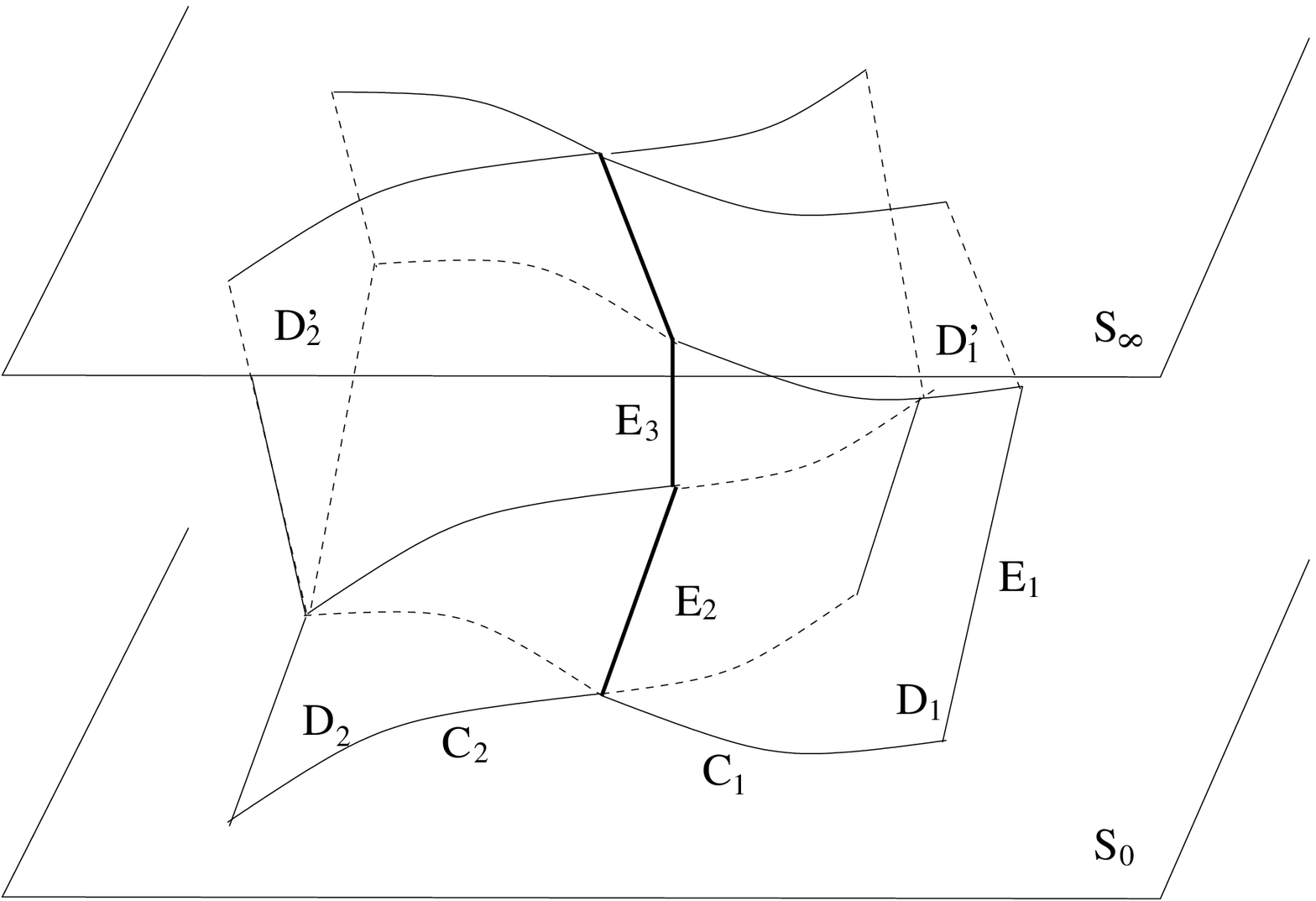}}

\noindent
Note that when performing the second blow-up along $C_2$, the 
exceptional divisor $D_1$ undergoes an embedded blow-up at the 
point $P$. This results in a reducible 
fiber with three components, the third component $E_3$ being the 
difference $E_1-E_2$. The strict transform $D_2^\pr$ also acquires 
a reducible fiber with a component identified with $E_3$. Therefore
we have $D_1\cdot D_2^\pr = E_3$. Note that $E_3$ is then a $(-1)$ 
curve in both surfaces, hence we can compute its normal bundle 
\eqn\normA{
N_{E_3/{\bar {B_3}}}\simeq \co(-1)\oplus \co (-1).}
If the blow-ups are performed in different order, the roles of $D_1, D_2$ 
are interchanged, so that that the extra component $E_3$ lies now in 
$D_2, D_1^\pr$. In this case, $E_3=E_2-E_1$. In fact the two models 
are related by a flop transition whose physical interpretation will 
be investigated below. 

At classical level, the map between heterotic and F-theory parameters 
is similar to the six dimensional case. The sizes of $E_1, E_2$ 
give the positions $\phi_1,\phi_2$ of the fivebranes in the M-theory 
interval, and are 
therefore generically different. Furthermore, the size of $E_3$ is 
related to the distance $|\phi_1-\phi_2|$ between the two fivebranes 
along the M-theory 
interval. In particular, note 
that as long as $E_3$ is of finite size, the two fivebranes do not really
intersect each other. Finally, as noted above, there are two choices for 
the curve $E_3$, namely $E_1 - E_2$ or $E_2 - E_1$. The process of going 
from one situation to the other can be interpreted as moving the two 
fivebranes past each other along the M-theory interval, with the 
fivebranes intersecting when
$E_1 = E_2$, \ie\ when $E_3$ shrinks to zero size. 

Thus, we find that intersecting
fivebranes on the heterotic side correspond to an 
isolated singularity in the base of the F-theory fourfold. 
If both curves are rational as stated above, the singularity in $B_3$ 
can be very
simply identified as a conifold by looking at the toric diagram of the 
fan of
the fourfold base.

\ifig\Conifold{Toric diagram of a conifold. The figure on the left gives 
the fan of $(P^1)^3$ blown up along two $P^1$'s in the base $(P^1)^2$. 
The fan of the base $(P^1)^2$ is described by the rays $R3,\ldots ,R6$. 
The conifold
singularity is described by the rays $R3,R6,R7$ and $R8$. The two small
resolutions of the conifold correspond to different triangulations of
the fan, and are shown on the right.}
{\epsfxsize4in\epsfbox{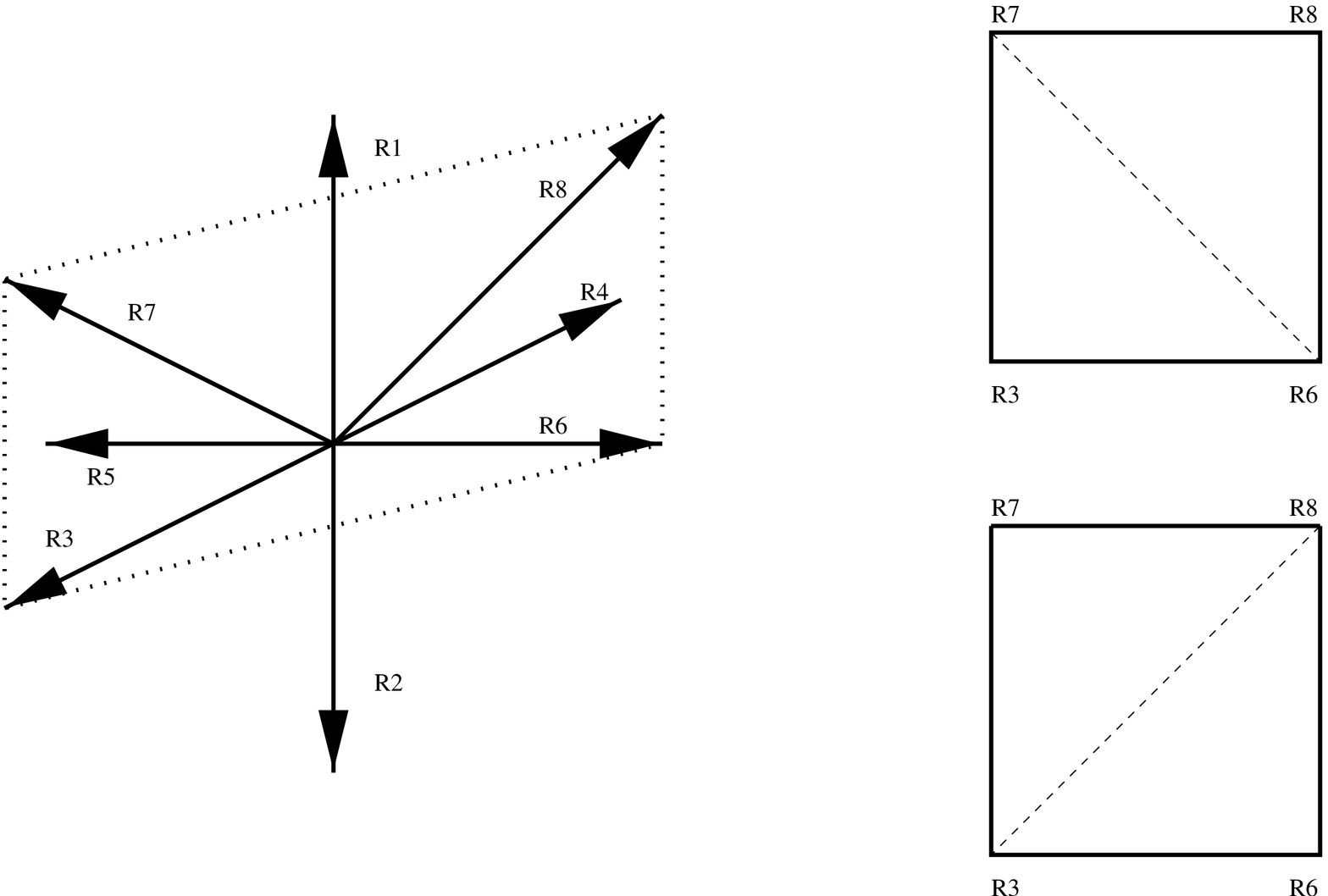}}

The two resolutions of the singularity are
then different
triangulations of the fourfold, related by a flop, which is precisely the
transition of $E_3$ from $E_1 - E_2$ to $E_2 - E_1$. It is
worth emphasizing
here that the conifold singularity is not in the \cyf\ itself, but in the
base ${\bar B}_3$. 
Alternatively, the normal bundle computation \normA\ 
shows that the singularity resulting from shrinking $E_3$ is a simple 
node. Thus,
separating the fivebranes on the M-theory interval, which gives $E_3$ a
non-zero volume, corresponds to resolving the conifold singularity, 
and the
two ways of moving them apart correspond to the two possible 
small resolutions of the conifold.

The relation between intersecting fivebranes and the conifold 
singularity has
been well studied recently
\nref\KW{I. Klebanov and E. Witten,''Superconformal Field Theory on 
Threebranes at a Calabi-Yau Singularity'', Nucl. Phys. {\bf B536} 
(1998) 199, hep-th/9807080.}%
\nref\MP{D. Morrison, M.R. Plesser, ``Non-Spherical Horizons, I'', 
hep-th/9810201.}%
\nref\OT{K. Oh and R. Tatar, "Three Dimensional SCFT from M2 Branes at
Conifold Singularities", JHEP {\bf 02} (1999) 025, hep-th/9810244.}%
\nref\AU{A. Uranga, ``Brane Configurations for Branes at Conifolds'',
hep-th/9811004.}%
\nref\DM{K. Dasgupta and S. Mukhi, ``Brane Constructions, Conifolds 
and M-Theory'', hep-th/9811139.}%
\nref\KOT{R. de Mello Koch, K. Oh and R. Tatar, "Moduli Space for Conifolds as 
Intersection of Orthogonal D6 branes", hep-th/9812097.}%
\refs{\KW , \MP , \OT , \AU, \DM , \KOT}. It has been shown there that
starting from a
configuration of two infinite Type IIA NS-fivebranes stretching along
$(012345)$ and $(012389)$, and
T-dualizing along $x_6$, we obtain Type IIB on a conifold. The
separation
of the NS-fivebranes along $x_6$ maps to a component of the $B$ field 
under the
T-duality. Furthermore, if we have fourbranes along $(01236)$, these 
map to
threebranes on the conifold singularity. In our case, we have 
fivebranes in
the heterotic theory. Thus we can heuristically relate our situation 
with the
IIA case by first going to M-theory by taking the strong coupling limit of
the IIA configuration (in which case the fourbranes now become fivebranes
along $(01236\&10)$), and then compactifying $x_7$ to an interval
($S^1 / Z_2$). We emphasize that this is only heuristic, since the 
analysis of
Refs. \refs{\KW , \OT, \AU, \DM, \KOT} is valid only for infinite branes,
while the
fivebranes here are compact (at least along the wrapped curves). 
However, since
the geometric description of the heterotic/F-theory vacua above is 
strictly
valid only when both the volumes of the section $\sigma$ (and hence the
wrapped curves) and the elliptic fiber are both large, we have reason 
to trust
our conclusions. 
In particular,
we see that in the F-theory case, we cannot turn on $B$ fields 
(as they are not
$SL(2,Z)$ invariant), so this is analogous to having $x_6=0$ in the 
IIA case.
Moreover, we
see that there is a very suggestive relation between the vertical 
heterotic
fivebranes and the IIA fourbranes, since in the M-theory limit, the
fourbranes become fivebranes wrapping $(6\&10)$, which form a torus (since
both $x_6$ and $x_{10}$ are compactified on circles), and the
vertical heterotic fivebranes wrap the elliptic fibre, which is also a 
torus.
Furthermore, in both cases, they are related to threebranes on the dual. 

We can use the analogy developed above to describe the situation when the
F-theory threebranes coincide with the conifold singularity in the base.
Note that the threebranes only move along ${\bar B}_3$, or rather, a 
section of 
the 
elliptic \cyf\ $X_4$. This follows from viewing the F-theory vacuum as 
Type
IIB compactified on ${\bar B}_3$, since the threebranes and sevenbranes 
are, 
strictly
speaking, Type IIB objects. When $n$ threebranes sit on the node of the
conifold (recall that it is really the base ${\bar B}_3$ which develops 
the  
conifold
singularity), the heterotic dual consists of intersecting horizontal
fivebranes with $n$ vertical fivebranes at the point of intersection. 
Thus, we
are again in a situation similar to the one discussed in 
\refs{\KW , \OT, \AU, \DM, \KOT},
where the M-theory dual of $n$ threebranes on a conifold consists of $n$
fivebranes along $(01236\&10)$ between two transverse fivebranes along
$(012345)$ and $(012389)$, respectively. However, since we 
are really in F-theory, we must switch off the the $B$ and $\tilde{B}$ 
fields.
This corresponds in the M-theory picture to setting the $x_6$ and $x_{10}$
separations of the transverse fivebranes to zero, so that there really 
is only
one set of fivebranes between them, and not two. 
Thus while the gauge 
theory in
the more general situation of \refs{\KW , \OT, \AU, \DM, \KOT} is an 
$SU(n)\times SU(n)$
theory, we see that in our case, we only get an $SU(n)$ gauge theory at 
the singularity. Note that unlike the situation in section 3.2, the 
arguments of \KOY\ showing the existence of an interacting theory 
on the fivebrane intersection are supported by the F-theory picture. 

The relation between the present situation and the one discussed in
\refs{\KW , \OT, \AU, \DM, \KOT} suggests a useful interpretation of the
resolution 
of the
conifold singularity in terms of the fivebranes. We note that the size of
$E_3$, the resolving $P^1$, is related to the separation the fivebranes 
along
the M-theory interval. We have argued before that this corresponds to the
$x_7$ direction in the picture
of \refs{\KW , \OT, \AU, \DM, \KOT}. Therefore, resolving the conifold should
correspond to separating the transverse fivebranes along $x_7$ in the
brane picture. In fact, separating the fivebranes along $x_7$ 
corresponds to
turning on a Fayet-Iliopoulos $D$-term in the Lagrangian of the 
$N=1$ $U(1)$
gauge theory describing the conifold, whose coefficient $\zeta$ 
is the $x_7$
separation of the branes 
\nref\GK{A. Giveon and D. Kutasov, ``Brane Dynamics and Gauge Theory'',
hep-th/9802067.}\refs{\KW, \GK}.
But, as shown in \KW, setting $\zeta \neq 0$ gives a resolution of the 
conifold singularity, and in fact there are two different resolutions
depending on the sign of $\zeta$. Moving the fivebranes past each other 
in the
$x_7$ direction thus corresponds to a flop transition in the dual theory,
exactly as in the heterotic picture above.

The conifold has another interesting property: it can be deformed to a 
smooth
manifold. Thus, we expect to be able to deform the base ${\bar B}_3$ away from 
the 
conifold. What is the heterotic dual of this deformation?

It was shown in 
\ref\GubK{S. Gubser, I. Klebanov, ``Baryons and Domain Walls in an 
N = 1 Superconformal Gauge Theory'', Phys. Rev. {\bf D58} (1998) 125025,
hep-th/9808075.} that the conifold metric
is deformed to a smooth metric when threebranes sit on the node of the
conifold. We will show in our case that even in the absence of the 
threebranes,
there are often deformations that smooth out the singularity.

Consider the
following example, first discussed in \refs{\GR}. We consider the 
heterotic
theory on the \cyt\ elliptically fibered over the Hirzebruch surface 
$F_1$, with
$E_8$ bundles chosen so that the dual \cyf\ is elliptically fibered over
$F_{100} = F_1\times P^1$. Now let us consider two horizontal fivebranes,
one wrapping the zero section $C_0$ and the other wrapping the fiber 
class $f$ of 
$F_1$. Since $C_0\cdot f=1$, these fivebranes intersect exactly once,
therefore we are in the situation discussed above.
Setting the size of the exceptional component $E_3$ 
to zero corresponds to moving the fivebranes on top of each other in the
M-theory interval so that they actually intersect, and we obtain a 
conifold
singularity in ${\bar B}_3$. However, we can now deform away from 
the conifold 
as follows. The infinity section $C_\infty$ of $F_1$ is in the same 
divisor class
as $C_0 + f$, but has one additional deformation modulus ($C_0$ has no
moduli as it has negative self-intersection and so cannot be moved, 
$f$ has exactly one modulus, and $C_\infty$ has two moduli). Thus, 
we can deform the
intersecting fivebranes into a single one wrapping $C_\infty$. 
On the F-theory
dual, we obtain a \cyf\ with different Hodge numbers 
(since the moduli are
different) but the same Euler number 
(since the number of threebranes is the
same).

On the other hand, if the heterotic base were the Hirzebruch surface $F_2$
instead, then we cannot deform the two curves $C_0$ and $f$ into a single
curve. Thus the node at the point of intersection of $C_0$ and $f$ 
cannot be
deformed. We see therefore that there is a global obstruction to the
deformation of the conifold singularity obtained by the intersection of 
the
two curves. This global obstruction can be simply stated as follows. The
number of moduli of a curve $C$ in the heterotic base $B_2$ is given by 
$h^0(N_{C/B_2})=h^0(\co_C(C))$, the number of deformations of the
normal bundle $N_{C/B_2}=\co_C(C)$ of $C$ in $B_2$. When $h^{1,0}(B_2)=
h^1(B_2, \co) = 0$, the number of moduli is also the 
dimension of the linear system of the divisor associated with the curve 
$C$, which is simply $h^0(B_2, {\cal O}(C))-1$.
Clearly, if we have two curves $C_1$ and $C_2$, then
we can deform to a single curve only if $C_1 + C_2$ has more moduli than
$C_1$ and $C_2$, so that the deformation exists only when
\eqn\defcond{h^0(\co_{(C_1+C_2)}(C_1+C_2)) > h^0(\co_{C_1}(C_1)) +
h^0(\co_{C_2}(C_2)).}

To summarize, we have found that intersecting horizontal fivebranes
are dual to isolated conifold singularities in the F-theory base. 
At classical level, the theory exhibits two branches corresponding to 
the small resolution and respectively to the deformation of 
singularities. The passage from one branch to the other represents an 
extremal transition in the fourfold geometry. Although the situation 
seems similar to the conifold singularities encountered in $N=2$ 
string vacua 
\nref\Str{A. Strominger, ``Massless Black Holes and Conifolds in 
String Theory'', Nucl. Phys. {\bf B451} (1995) 96, hep-th/9504090.}%
\nref\GMS{B. Greene, D.R. Morrison, A. Strominger, ``Black Hole 
Condensation and the Unification of String Vacua'', Nucl. Phys. 
{\bf B451} (1995), 109, hep-th/9504145.}%
\refs{\Str,\GMS}, there are important differences. 
First note that the $S^3$ cycle present in the local deformation of the 
conifold is in general homologically trivial in the deformed F-theory 
base. If $C_1, C_2$ are rational curves, this follows easily by 
noting that the third Betti number of the deformed base is zero. 
Therefore, the usual restrictions on K\"ahler resolutions imposed by 
the presence of three-cycles are absent in this case. In particular, 
small K\"ahler resolutions of single isolated conifold are allowed.

Physically, this means that on the deformation branch, we cannot 
identify a state of vanishing mass at the singularity. Moreover, 
even if $S^3$ were homologically nontrivial, a wrapped threebrane would 
still not define a stable BPS state due to the reduced amount of 
supersymmetry. At the same time, a threebrane wrapped around the 
$S^2$ cycle of a small K\"ahler resolution will give rise to a 
stable BPS string whose tension goes to zero as we approach the 
singularity. In the context of Calabi-Yau conifolds, this string has 
been identified with a flux tube between charged sources in the confining
phase of an $N=2$ gauge theory 
\ref\GMV{B. Greene, D.R. Morrison, C. Vafa, ``A Geometric Realization 
of Confinement'', Nucl. Phys. {\bf B481} (1996) 51, hep-th/9608039.}. 
Here, single conifold transitions are allowed, therefore such an 
interpretation is no longer valid. Instead, we can regard the stable 
tensionless string as (weak) evidence for the existence of an interacting 
superconformal theory localized at the singularity. This would be 
dual to the theory on intersecting heterotic fivebranes. 

Thus far, we have only considered curves of genus zero, \ie\ $P^1$'s. 
Let us
now see how this discussion needs to be modified when we have curves of
higher genus. First, according to section 2, the F-theory dual of a 
fivebrane wrapping a
curve of genus $g$ is once again a blowup of $B_3$ over the same curve, 
the size of the blowup giving the position of the fivebrane in the
M-theory interval. When two such curves intersect in $n$ points, the 
local geometry near each intersection point is identical to that
represented in Fig.2. Therefore we obtain $n$ exceptional $P^1$ 
components of type $E_3$ which are in the same cohomology class,
either $E_1 - E_2$ or $E_2 - E_1$. Since the total space is K\"ahler, 
they are also of the same size
giving the separation between the fivebranes on the interval. When 
the two fivebranes coincide, all the exceptional $P^1$ components 
shrink simultaneously and we obtain $n$ isolated conifold singularities. 
This picture shows that the $n$ isolated singularities cannot be 
resolved independently. There are
precisely two ways of resolving the singularity 
(which are related by a flop), corresponding to the two ways of 
separating the fivebranes on the interval. Once again, deforming the 
two curves $C_1$ and $C_2$ into a
single curve in the class $C_1 + C_2$ is a deformation away from the
singularity. The global obstruction to this is again given by 
Eqn.\defcond. 

\subsec{Parallel Fivebranes}

Now consider the situation when we have two horizontal fivebranes 
wrapping the
same curve $C$. Again, we first consider the case when the curve in 
question
is a $P^1$. On the F-theory side, the base ${\bar B}_3$ now has two 
intersecting
exceptional $P^1$'s $E_1, E_2$ fibered over $C$ ($E_3$ is the proper 
transform of the original fiber)

\ifig\Multiple{The blow-up geometry corresponding to two fivebranes
wrapping the curve $C$.}{\epsfxsize3.0in\epsfbox{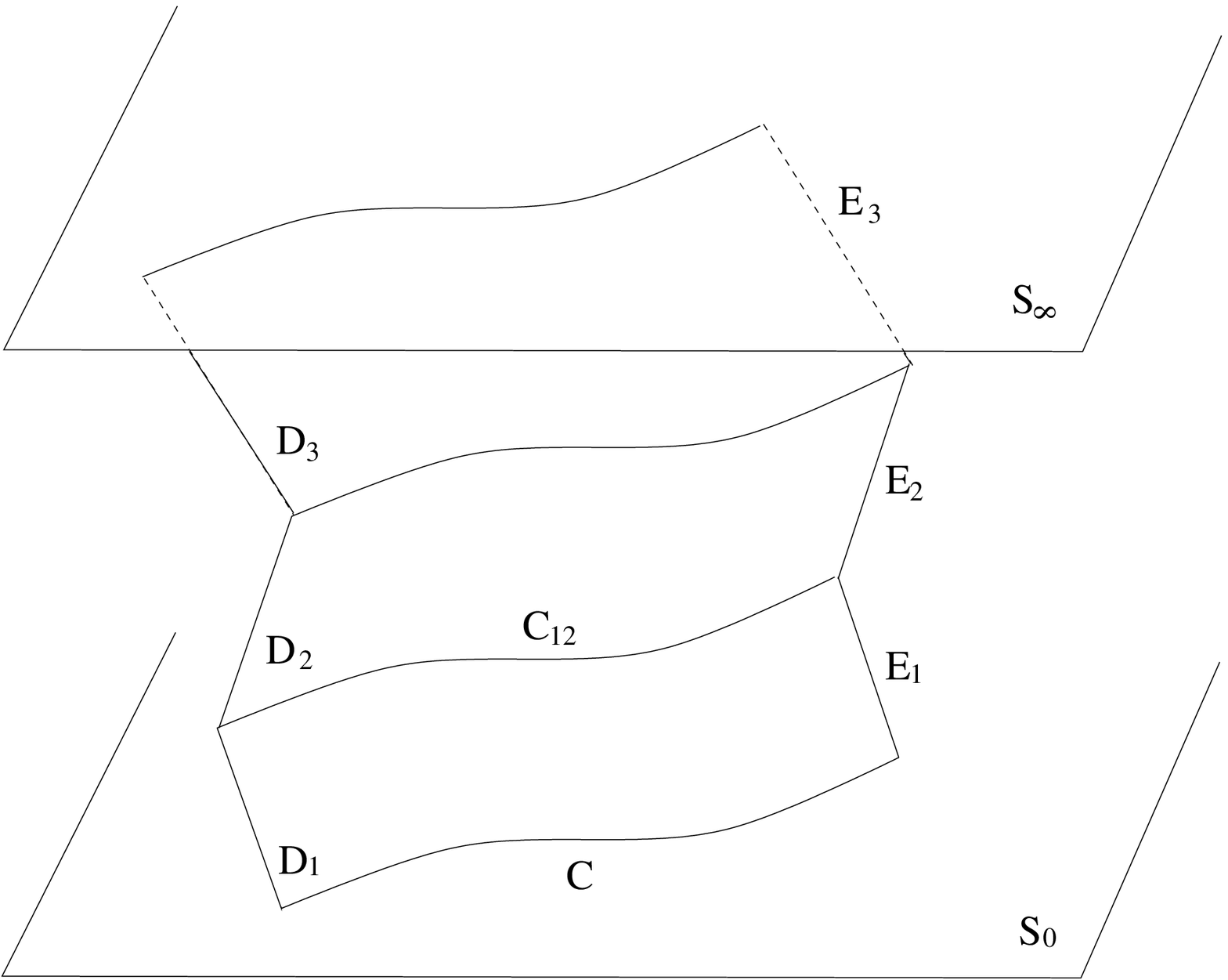}}

\noindent
with the size of the $E_2$ giving the 
mutual separation of the fivebranes along the M-theory interval. 
Blowing down this
$P^1$, therefore, corresponds to moving the fivebranes on top of each 
other.
This gives an $A_1$ singularity over $C$ in ${\bar B}_3$. More generally, if 
we have
$k$ coincident fivebranes wrapping $C$, we would get an $A_{k-1}$ 
singularity
over $C$ in ${\bar B}_3$. In addition, if we have $n$ coincident vertical 
fivebranes,
this corresponds on the F-theory side to $n$ threebranes on the $A_{k-1}$
singularity. As before, we see that this is very similar to the situation
discussed in Refs. \refs{\DM, \AU}, except that the $B$ fields are zero in
our situation, corresponding to zero separation of the branes along 
$x_6$ and
$x_{10}$ in the brane description. For clarity, we will restrict our 
attention
to the case when we have only two parallel fivebranes for the rest of 
this
section. The generalization to more branes is elementary.

Apart from the obvious resolution of this singularity (blowing up the $P^1$
again, corresponding to separating the fivebranes), we can deform away 
from
it as follows. First, if $C$ is movable
(\ie\ if $h^0(\co_C(C)) > 0$), then we can deform away from the
singularity by moving one curve away from the other in $B_2$. If $C.C=0$,
then the two curves no longer intersect, and the corresponding 
${\bar B}_3$ is
smooth. However, if $C\cdot C >0$ ($C\cdot C < 0$ is impossible here, since
we have assumed that $C$ is movable), then the curves intersect in a set of 
points,
and ${\bar B}_3$ becomes a conifold, as in the previous section, with a 
set of
isolated nodes. We can then either resolve the conifold by blowing 
it up, \ie\
separating the fivebranes, or by deforming it further. This deformation is
only possible if $2C$ has a smooth section different from $C$, \ie\ if
$h^0(\co_{2C}(2C)) > 2 h^0(\co_C(C))$. In this case, 
the curves
can be deformed into a single curve, and we get a single fivebrane 
wrapping
a smooth curve in the class $2C$. Of course, in this case, we could have
directly deformed the two curves into a single curve without passing 
through
the conifold phase. 
If $C$ is not movable, then we cannot deform to a conifold. However, if 
$h^0(\co_{2C}(2C)) > 2 h^0(\co_C(C))$, then we can still
deform to a single curve, because $2C$ is then movable. If, however, 
$2C$ is also not movable, then we cannot deform away at all.
The geometrical analysis can be easily generalized to curves of higher 
genus.

Note that in contrast with the previous cases, the low energy 
theory obtained here is essentially associated to a curve of 
$A_{k-1}$ singularity in Type IIB theory. Therefore we will have 
an interacting theory which can be described as the compactification 
of the $(2,0)$ field theory on a Riemann surface. If vertical fivebranes 
are added to the picture, the dual F-theory phenomenon consists 
of threebranes transverse to a curve of $A_{k-1}$ singularities 
in IIB theory. This is a familiar situation encountered many 
times in the literature, starting with 
\ref\DM{M. Douglas and G. Moore, ``D-branes, Quivers, and ALE 
Instantons'', hep-th/9603167.}. However, the resulting effective 
theory may present some complications due to the finite size 
of the singular curve. This will be discussed in the next subsection. 

Finally, we can combine the situation here with that of the previous
subsection, and consider $k_1$ horizontal fivebranes wrapping $C_1$, 
$k_2$
horizontal fivebranes wrapping $C_2$, and $n$ vertical fivebranes at 
a point
of intersection of $C_1$ and $C_2$. First, ignoring the vertical 
fivebranes, the dual F-theory degeneration consists of two 
curves of $A_{k_1-1}$ and $A_{k_2-1}$ singularities intersecting 
transversely in ${\bar B}_3$. This results in a nonabelian conifold 
singularity 
\nref\DD{M. Bershadsky, V. Sadov, C. Vafa, ``D-strings on D-manifolds'',
Nucl.Phys. {\bf B463} (1996) 398, hep-th/9510225.}%
\nref\HOV{K. Hori, H. Ooguri, C. Vafa, ``Non-Abelian Conifold 
Transitions and N=4 Dualities in Three Dimensions'', Nucl. Phys. 
{\bf B504} (1997) 147, hep-th/9705220.}%
\refs{\DD,\HOV}.
Near each intersection point, 
this is again similar to a situation considered in Ref. \refs{\AU}, 
where the
singularity was shown to be of the form $xy=z^{k_1}w^{k_2}$. Note again 
that
the $x_6$ separation of the branes is zero since we have to set the $B$
fields to zero. 
It is an easy exercise to describe geometrically the 
various
deformations and resolutions of this situation. The specific example of 
$k_1=1, k_2=2$ was worked out in \S{4} of Ref. \refs{\AU}.
Bringing the $k$ vertical fivebranes near an intersection point
corresponds to placing $n$ threebranes at a nonabelian conifold 
singularity in F-theory. 
The description of the low energy effective theory is quite 
difficult in this case, as will be detailed in the following. 

\subsec{Merging Horizontal and Vertical Fivebranes -- A Second Puzzle}

Consider the situation discussed in the previous subsection, namely 
$n$ vertical fivebranes intersecting $k>1$ horizontal fivebranes
wrapped around a curve $C$ in $B_2$. 
Here we encounter a qualitatively new phenomenon. The 
fivebranes can merge together into a single fivebrane wrapping a 
smooth curve $C^\pr$ in the Calabi-Yau space $Z$. An obvious necessary
condition for this is $(kC+nf)\cdot C>0$, \ie\ $n > -k C\cdot C$.
The resulting fivebrane 
is, in a sense, ``skew''. The case $k=1$ was discussed in section 3.2. 

Therefore, it follows that there is a new branch of the theory along 
which the low energy effective action consists of $g(C^\pr)$ abelian 
gauge fields whose couplings are governed by the Jacobian of $C^\pr$. 
The main problem is to understand the F-theory origin of this new 
branch. As a first step, note that this phenomenon cannot admit a 
pure geometrical interpretation. This is because horizontal and 
vertical fivebranes map to very
different objects in F-theory. The horizontal fivebranes are described 
in terms
of blowup modes in the base, which can be described purely in terms 
of the
fourfold geometry, but the vertical fivebranes map to threebranes, 
which are
not related in any way to the geometry - rather, they constitute 
additional
non-geometric data necessary for the complete description of the F-theory
vacuum. Therefore, when these two kinds of fivebranes merge into a single
fivebrane, it is unlikely to find a pure geometric description. 

In the following, we suggest a possible resolution of this puzzle based 
on a careful analysis of threebranes transverse to a curve of $A_{k-1}$ 
singularities. In principle, threebranes transverse to an $A_{k-1}$ 
singularity are described by an $A_{k-1}$ quiver gauge theory 
\DM. The space time parameters -- blow-up modes and theta angles --
are realized as FI terms in the brane gauge theory. However, at the 
same, the spacetime moduli are coordinates along the flat directions 
of the $A_{k-1}$ (2,0) theory localized at the singularity. 
A similar situation is encountered in the $(1,0)$ theories with 
tensor multiplets discussed in 
\nref\Int{K. Intriligator, ``RG Fixed Points in Six Dimensions via 
Branes at Orbifold Singularities'', Nucl. Phys. {\bf B496} (1997) 177,
hep-th/9702038.}%
\nref\BI{J.D. Blum, K. Intriligator, ``New Phases of String Theory and 
6d RG Fixed Points via Branes at Orbifold Singularities'', Nucl. Phys. 
{\bf B506} (1997) 223, hep-th/9705044.}%
\refs{\Int,\BI}.
In the 
present situation, the spacetime $(2,0)$ is compactified on a Riemann 
surface of finite size, therefore it yields an interacting 
four dimensional theory with no Lagrangian description. 
Moreover, the $(2,0)$ degrees of freedom must interact nontrivially with 
the threebrane degrees of freedom, as a result of the previous 
interplay between space time moduli and brane gauge theory.
Therefore, we conclude that the theory associated to F-theory 
threebranes transverse to a curve of $A_{k-1}$ singularities
must be a complicated interacting fixed point. The new fivebrane 
branch found above can then be interpreted as a low energy 
Coulomb branch emerging from this fixed point. In particular, it is 
a non-geometric branch.

When $C\cdot C = -2$, $\Sigma = \pi^*{C} \simeq P^1\times T^2$, and
there is a brane construction for the above situation. The above
Coulomb branch can then be interpreted in terms of fractional
branes \ref{\Ang}{A. M. Uranga, private communication.}. It is
not clear if this description holds in the more general cases.

Another possibility \ref{\Ed}{E. Witten, private communication.}, is to
compactify on a further circle to M-theory.
In this picture, the F-theory threebranes map to membranes. Deforming 
$kC+nf$ to a single irreducible curve might correspond to absorbing $n$ 
membranes and turning on a four-form field strength flux carrying 
$n$ units of membrane charge. Lifting back to F-theory, this would 
map to turning on $H,\tilde H$ fluxes. We are currently investigating
this possibility and hope to report on it in the future. 

Finally, note that the situation is even more complicated when vertical 
fivebranes sit at a point of intersection of multiple horizontal 
fivebranes. 

\newsec{Discussion of the Superpotential}

In the previous sections, we have studied the $N=1$ moduli space 
from an essentially classical point of view. Here we will discuss
how the classical picture is modified by nonperturbative instanton 
effects. From the heterotic perspective, there can be various 
nonperturbative phenomena generated by spacetime or worldsheet 
instantons. All these effects have a simple and unitary F-theory 
interpretation in terms of threebrane instantons wrapping 
divisors in the threefold base. 
In the following we will give a systematic treatment of these effects 
for the small instanton degenerations considered so far. Note that 
in order obtain correct results, the zero mode computations must be 
performed away from the zero degeneration limit, that is 
for a smooth elliptic Calabi-Yau fourfold $\pi:X\rightarrow B_3$.
This can be seen, for example, by taking into account the M-theory 
origin of the instantons explained in \W.
In order to obtain such a smooth model, the base $\bar B_3$ must be 
blown-up a number of times along certain curves contained in the 
two $II^*$ sections. As before, let ${\bar B}_3$ denote the blown-up base.
Many of the results of this section have been obtained independently by
A. Grassi \ref{\AGr}{A. Grassi, to appear.}. We are grateful to her for
sharing her results with us.

{\it i) Heterotic worldsheet instantons.} Generically, these
correspond to threebranes wrapping divisors 
$W\subset {\bar B}_3$ which are 
vertical with respect to the map $p:{\bar B}_3\rightarrow B_2$ \W.\ Here, 
this is still valid if $W$ is the pull back of a generic curve 
$C$ in the base, other than the support of the exceptional locus. 
In this case, we can apply directly the results of \G.\ 
The number of zero 
modes is given by 
\eqn\arthgenA{
\chi\left(\co_{\pi^*W}\right)={1\over 2}K_{{\bar B}_3}W^2.}
Since $W=p^*C$, $W^3=0$ and the adjunction formula 
\eqn\adjunct{
K_W=\left(K_{{\bar B}_3}+W\right)\cdot W}
shows that 
\eqn\arthgenB{
K_{{\bar B}_3}W^2=K_W\cdot W=-2(C^2)_{B_2}.}
Therefore, only divisors supported on curves with $(C^2)_{B_2}=-1$
can contribute. As the base $B_2\simeq F_n,\ 0\leq n\leq 2$, it follows
that a superpotential can be generated only if $n=1$.
The analysis of \refs{\W,\G} shows that in this case, a superpotential is 
actually generated.

{\it ii) Noncritical string instantons.} Regarding the heterotic theory 
as M-theory on an interval, there are two types of noncritical strings 
corresponding to membranes stretching between two fivebranes or 
membranes stretching between a fivebrane and a nine dimensional wall.
Some effects associated with these instantons have been discussed in 
\nref\M{P. Mayr, ``Mirror Symmetry, N=1 Superpotentials and 
Tensionless Strings on Calabi-Yau Four-Folds'', Nucl. Phys. 
{\bf B494} (1997) 489, hep-th/9610162.}%
\nref\KS{S. Kachru, E. Silverstein, ``Chirality Changing Phase 
Transitions in 4d String Vacua'', Nucl. Phys. {\bf B504} (1997) 272, 
hep-th/9704185.}%
\refs{\M,\KS}. 
These are BPS strings whose tension is proportional to the separation 
between the branes. Therefore, taking into account the moduli map 
developed in the previous sections, they can be naturally identified 
with Euclidean threebranes wrapping the exceptional divisors in F-theory. 
More precisely, let us consider a configuration of $k$ fivebranes 
wrapping a curve $C$ in $B_2$ and separated along the interval. 
The corresponding F-theory geometry consists of a sequence of
divisors $D_1,D_2\ldots D_k, D_{k+1}$ with normal crossings 
as explained in section 4.2. Threebranes wrapped on the divisors 
$D_1, D_{k+1}$ correspond to membranes stretching between the first and
the last fivebrane and the nine dimensional walls respectively. 
At the same time, threebranes wrapped on $D_1\ldots D_k$ correspond to
membranes stretching between consecutive fivebranes. The number of zero 
modes can be computed again using formula \arthgenA.\ In fact a simple 
local computation shows that the fibers $E_1, E_{k+1}$ of $D_1, D_{k+1}$ 
are negative extremal rays in the blown-up threefold base, therefore 
we can apply the results of \G\ (Ex. 2.9.1 and Prop. 3.4). This shows 
that 
\eqn\arthgenD{
\chi\left(\co_{\pi^*D_1}\right)=
\chi\left(\co_{\pi^*D_{k+1}}\right)=1-g(C)}
where $g(C)$ is the genus of the curve. Moreover, it can be shown 
as in \G\ that the Hodge numbers of these divisors are 
\eqn\hodgeA{
h^{0,0}=1,\qquad h^{0,1}=g(C),\qquad h^{0,2}=h^{0,3}=0.}
Therefore the divisors $D_1, 
D_{k+1}$ contribute if and only if $C$ is a rational curve. 

This formula does not apply to the middle divisors $D_1\ldots D_k$ since 
their fibers are not negative extremal rays. The number of zero modes 
can be computed recursively as follows. For simplicity consider the case 
$k=2$ represented in \Multiple.\ Let $n_i$, $i=1,2,3$ denote the degrees of
the rulings of the three
divisors. Note that by construction, 
\eqn\degrel{
n_1=n_2+(C^2)_{B_2}}
and the divisors $D_1, D_2$ intersect along a common section $C_{12}$
isomorphic to $C$.
The adjunction formula \adjunct\ shows that 
\eqn\arthgenE{
K_{{\bar B}_3}D_2^2=\left(K_{D_2}-D_2^2\right)\cdot D_2}
where 
\eqn\canclass{
K_{D_2}=-2C_{12}+(2g(C)-2-n_2)E_2.}
Furthermore, a local computation shows that the restriction of the 
normal bundle $N_{D_2/{\bar B}_3}$ to the $P^1$ fiber $E_2$ is of degree 
$-2$. Therefore,
we have 
\eqn\normbd{
N_{D_2/{\bar B}_3}=-2C_{12}+aE_2}
where $a$ is an integer number which can be determined as follows. 
The triple intersection $D_1D_2^2$ can be computed in two equivalent ways
\eqn\intrel{\eqalign{
D_1D_2^2 &= (C_{12}^2)_{D_1}=n_1\cr
&= C_{12}\cdot D_2=2n_2+a.\cr}}
Using also \degrel,\ we find
\eqn\coeff{
a=(C^2)_{B_2}-n_2.}
Finally, \arthgenE,\ \canclass\ and \normbd\ imply that 
\eqn\arthgenF{
\chi\left(\co_{\pi^*D_2}\right)=-(C^2)_{B_2}+2g(C)-2=-C\cdot K_{B_2}.}
It is clear that this recursive step can be applied for each of 
the divisors $D_2\ldots D_k$ yielding the same result. 
The Hodge numbers can be computed using the general expressions in
\G\ and the Riemann-Roch theorem on the $D_i$
\eqn\hodgeB{
h^{0,0}=1,\qquad h^{0,1}=g(C),\qquad h^{0,2}=-C\cdot K_{B_2}+g(C)-1,
\qquad h^{0,3}=0.}
Note that again there is an exceptional case, namely 
$B\simeq F_2$, $C=C_0$ when the above general formulae do not hold. 
In that case, we obtain similarly 
\eqn\hodgeC{
h^{0,0}=1, \qquad h^{0,1}=1,\qquad h^{0,2}=h^{0,3}=0.}
Therefore, although $-C\cdot K_{B_2}=1$ is a necessary condition for 
the generation of a superpotential, it may not be sufficient \W. 
A sufficient condition also requires $g(C)=0$, showing that the 
contribution is certain for rational curves satisfying 
$\left(C^2\right)_{B_2}=-1$. If the base is a Hirzebruch surface 
of degree $0\leq n\leq 2$, this can be realized only when $n=1$ and $C$ 
is a rational curve.

{\it iii) Exceptional Instantons.} In addition to the effects discussed so
far, there are also contributions 
coming from the resolution of the sections of $E_8$ 
singularities of the elliptic fibration. 
These have been shown in \refs{\KV,\CV} to give the expected 
gauge theoretic nonperturbative superpotentials and will not be 
discussed here.

\subsec{Toric Description}

It is interesting also to consider the
computation of the zero modes in the
light of toric geometry, which provides a method of explicitly 
constructing
\cyf s as hypersurfaces in toric varieties. Such a \cyf\ $X$ is 
described by
a dual pair of five dimensional reflexive polyhedra $(\Delta, \nabla)$. 
In particular, the points in $\nabla$ correspond to divisors in the \cym . 
(See,
for example, 
\ref\HS{H. Skarke, ``String Dualities and Toric Geometry: 
An Introduction'', hep-th/9806059.} for a useful review). For
any point $p$ in $\nabla$, the arithmetic genus of the corresponding 
divisor
in the \cyf\ is given by a formula of 
\ref\KLRY{A. Klemm, B. Lian, S-S. Roan
and S-T. Yau, ``Calabi-Yau fourfolds for M- and F-Theory 
compactifications'', Nucl. Phys. {\bf B518} (1998) 515 hep-th/9609239.}. 
If $p$ is interior to a face $\Theta ^*_p$ in
$\nabla$, and $\Theta _p$ is the dual face in $\Delta$ (defined by
$\Theta_p = \{q \in \Delta | <q,v> = -1, \forall  v \in \Theta^*_p\}$, 
note
that $dim(\Theta_p) + dim(\Theta^*_p) = 4$), and   
$l(\Theta)$ is the number of lattice points interior to $\Theta$, the formula
of \KLRY\ is
\eqn\toricarithgen{
\chi(D_p, {\cal O}(D_p)) = 1 - (-1)^{dim(\Theta_p)}l(\Theta_p),}
where $D_p$ is the divisor in $X$ corresponding to the point $p$. (If
$dim(\Theta^*_p)=4$, the point $p$ corresponds to a divisor in the embedding
variety that does not intersect the space $X$, and hence does not give
a divisor on $X$).

Since we are studying F-theory compactifications, $X$ is an elliptic
fibration, \ie\ there is a
projection
$\pi: X \rightarrow  B_3$ whose generic fiber is an elliptic curve ${\cal E}$, 
and the vertical divisors (which project down to divisors in $B_3$) are the
only ones that contribute to the superpotential. Now, by theorems in
\nref\AKMS{A. Avram, M. Kreuzer, M. Mandelberg and H. Skarke, ``Searching for
K3 fibrations'', Nucl. Phys. {\bf B}494 (1997) 567, hep-th/9805190.}%
\nref\KS{M. Kreuzer and H. Skarke, ``Calabi--Yau 4-folds and Toric
fibrations'', J. Geom. Phys. 26 (1998) 272, hep-th/9701175.}%
\refs{\AKMS, \KS}, 
the polyhedron $\nabla$
contains a slice $\nabla_{\cal E}$ through the origin, where $\nabla_{\cal E}$
is the polyhedron describing the elliptic fiber (which is a torus, and hence
Calabi--Yau) as a hypersurface in a toric variety. The theorem also assures
us that
there exists a projection acting on $\nabla$, which projects $\nabla_{\cal E}$ 
to a point, and whose image is the fan $\Sigma_{B_3}$ of $B_3$. Thus, points
in $\nabla$ project down to points in $\Sigma_{B_3}$, and hence describe
divisors in $B_3$. Furthermore, in order that a heterotic dual exist, $X$
must also admit a $K3$ fibration that is consistent with the elliptic
fibration structure. This implies that $B_3$ is itself a $P^1$ fibered over
$B_2$, the base of the heterotic threefold $Z$.
\ifig\Blowups{In (A), we have the fan of $F_{n,0,0}=F_n\times P^1$, the base
of the fourfold dual to the heterotic compactification on the elliptic
\cyt\ with base $F_n$. In (B),
we have the fan of $F_{n,0,0}$ blown up three times over the zero section $C_0$
of $F_n$, corresponding to three fivebranes wrapping $C_0$ in the heterotic
base $F_n$.}
{\epsfxsize5in\epsfbox{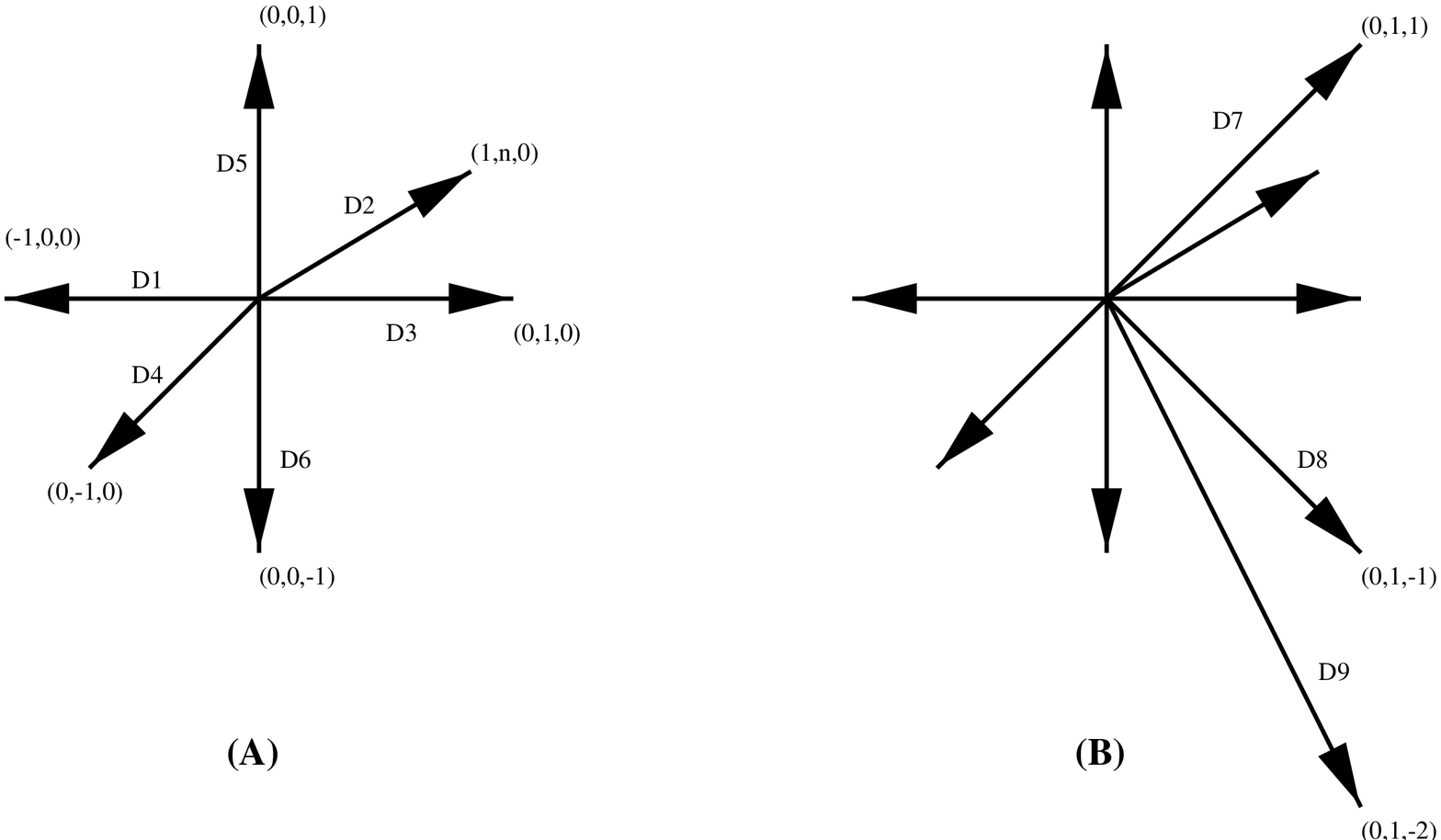}}
When $k$ fivebranes wrap a curve $C$ in $B_2$
which corresponds to a toric divisor, the corresponding base ${\bar B_3}$
acquires
$k$ blowup modes over the corresponding divisor (see \Blowups ).
Thus, we get
a line of points $p_1, p_2,\ldots , p_{k+1}$ all of which project down to the
divisor $C$ in $B_2$ under the projection $P: {\bar B_3} \rightarrow B_2$.
The fact that all of these points lie in a straight line indicates that they
all lie in
the same face of $\nabla$. In the simplest cases, $p_1$ and $p_{k+1}$ are
vertices in $\nabla$,
while $p_2,\ldots ,p_k$ lie in the edge joining $p_1$ and $p_{k+1}$. In
general, $p_2,\ldots ,p_k$ are interior to a face
$\Theta^*$ of $\nabla$, while $p_1$ and $p_{k+1}$ will usually
lie on the boundary of $\Theta^*$,
and hence in faces of lower dimension. 
Now the expression
\toricarithgen\ yields the same result for all points interior to a given
face $\Theta^*$ of $\nabla$. Thus we see immediately that $p_2,\ldots, p_k$
all have exactly the same arithmetic genus, and therefore, either they all
contribute to the superpotential or they all do not. Moreover, it follows that
the points $p_1$ and $p_{k+1}$ can have a different arithmetic genus.

We can actually compute the arithmetic genus of the divisors in the fourfold
that correspond to heterotic worldsheet instantons and noncritical string
instantons as follows. The points of $\nabla$ for the fourfold $X$ can be put
in the form
$(x,y,z,u,v)$, such that the slice $(0,0,0,u,v)$ gives $\nabla_{\cal E}$, and
the slice $(0,0,z,u,v)$ gives $\nabla_{K3}$.
The projection $P_3:(x,y,z,u,v)\rightarrow (x,y,z)$ yields the fan
$\Sigma_{B_3}$ of $B_3$, the base of the elliptic fibration, while
$P_2:(x,y,z,u,v)\rightarrow (x,y)$
gives the fan $\Sigma_{B_2}$ of the base $B_2$ of the $K3$ fibration.
The points $q \in \Delta$ satisfy $<q,p>\, \ge -1, \forall p \in \nabla$.
In terms of $\Delta$, the polyhedron $\Delta_{K3}$ of the $K3$ fiber is seen
as a projection $\Pi_{K3}:\{a,b,c,d,e\}\rightarrow\{c,d,e\}$, while
$\Delta_{\cal E}$ of the elliptic fiber is given by
$\Pi_{\cal E}:\{a,b,c,d,e\}\rightarrow\{d,e\}$\refs{\AKMS, \KS}. (We will use
round brackets
to denote points in $\nabla$, and curly brackets to denote points in
$\Delta$.)
For example, the polyhedron given by the points
$$\eqalign{&{\tt
(1,1,0,2,3), (0,1,0,2,3), (0,0,1,2,3), (0,0,0,2,3), (0,0,0,1,2), (0,0,0,1,1),
(0,0,0,0,1)}\cr
&{\tt (0,0,0,0,0), (0,0,0,0,-1), (0,0,0,-1,0), (-1,0,0,2,3), (0,-1,0,2,3),
(0,0,-1,2,3)}\cr}$$
describes a fourfold elliptically fibered over $F_1\times P^1$, which can also 
be viewed as a $K3$ fibration over $F_1$.

Consider first the case of the worldsheet instanton. This corresponds to a
point $p \in \nabla$ that 
projects down to a point ${\tilde p} \in \Sigma_{B_2}$ under $P_2$, which
describes a
divisor (curve) $C_{\tilde p}$ in $B_2$. Now
$C_{\tilde p}\cdot C_{\tilde p} \ge -2$, otherwise the elliptic
fibration will be singular over $C_{\tilde p}$. In such a case we do not have
heterotic
worldsheet instantons, but rather instantons associated with the exceptional
fibers. For $C_{\tilde p}\cdot C_{\tilde p} > -2$, the point $p$ is
a vertex,
whereas for $C_{\tilde p}\cdot C_{\tilde p} = -2$, $p$ is interior to an edge.
Now, the points $q$ which are ${\it interior}$ to a face $\Theta$ in
$\Delta$
satisfy $<q,v> = -1, \forall v \in \Theta^*$ and $<q,v>\, \ge 0, \forall v
\notin \Theta^* \, (v \in \nabla)$.
But $\nabla$ contains the reflexive polyhedron $\nabla_{K3}$ as a slice.
The only point in $\Delta_{K3}$ that has non-negative product with ${\it all}$
the points in $\nabla_{K3}$ is the origin $\{0,0,0\}$, since, by reflexivity,
it is the unique interior point of $\Delta_{K3}$. Since
$\Pi_{K3}(\Delta)=\Delta_{K3}$, it follows that the only points in $\Delta$ 
that have non-negative product with all
the points in $\nabla_{K3} \subset \nabla$ are of the form $\{a,b,0,0,0\}$.
Now, all the
points interior to $\Theta_p$ must have non-negative product with any point in
$\nabla$ not in $\Theta^*_p$. In particular, they have
non-negative product with any point in
$\nabla_{K3} \subset \nabla$, and so must be of the form $\{a,b,0,0,0\}$. It
is then a
simple matter to count $l(\Theta_p)$, the number of points interior to
$\Theta_p$.

For $C_{\tilde p}\cdot C_{\tilde p} > -2$, we find $l(\Theta) =
1 + C_{\tilde p}\cdot C_{\tilde p}$. Since $p$ is a vertex, 
$dim(\Theta_p)=4$, and \toricarithgen\ gives
$$\chi(D_p,{\cal O}(D_p))=-C_{\tilde p}\cdot C_{\tilde p}.$$
Using the results of \KLRY\ we also find $$h^{0,0}=1,\qquad h^{1,0}=0,\qquad
h^{2,0}=0,\qquad h^{3,0}=1+C_{\tilde p}\cdot C_{\tilde p},$$
so that a superpotential is generated
if and only if
$C_{\tilde p}\cdot C_{\tilde p}=-1$.

When $C_{\tilde p}\cdot C_{\tilde p}=-2$, we get $l(\Theta) = 1$. But
$dim(\Theta_p)=3$ (since $p$ must lie in an edge, otherwise the fourfold 
$X$ will not
admit a $K3$ fibration consistent with the elliptic fibration structure),
so \toricarithgen\ gives
$$\chi(D_p,{\cal O}(D_p))=2=-C_{\tilde p}\cdot C_{\tilde p}.$$ In fact, we find
$$h^{0,0}=1,\qquad h^{1,0}=0,\qquad 
h^{2,0}=1,\qquad h^{3,0}=0.$$ 

Now consider the case of the noncritical string instantons. We now have a line
of points $p_1,\ldots,p_{k+1} \in \nabla$ all of which project down to the same
point ${\tilde p} \in \Sigma_{B_2}$. Once again, $C_{\tilde p}\cdot
C_{\tilde p} \ge -2$, otherwise we have exceptional fibers. 

For $C_{\tilde p}\cdot C_{\tilde p} > -2$ the points
$p_j,\, 2 \le j \le k$ lie in
the edge
$\Theta^*_{1,k+1}$
joining $p_1$ and $p_{k+1}$. Once again, we see that all the points interior to
$\Theta_{1,k+1}$ are of the form $\{a,b,0,0,0\}$, and
$l(\Theta_{1,k+1})= 1 + C_{\tilde p}\cdot C_{\tilde p}$. Thus, for the points
$p_j,\, 2 \le j \le k$,
we get, using $dim(\Theta_{1,k+1})=3$ and \toricarithgen ,
$$\chi(D_{p_j},{\cal O}(D_{p_j}))=2+C_{\tilde p}\cdot C_{\tilde p}=
-K_{B_2}\cdot C_{\tilde p}.$$ We also find
$$h^{0,0}=1,\qquad h^{1,0}=0,\qquad
h^{2,0}=1+C_{\tilde p}\cdot C_{\tilde p}=-K_{B_2}\cdot C_{\tilde p}-1,\qquad
h^{3,0}=0.$$
But $\Theta_{1,k+1}$ is the common face of $\Theta_1$ and $\Theta_{k+1}$. We
see that all the points of the form $\{a,b,0,0,0\}$ in $\Theta_1$ lie in
$\Theta_{1,k+1}$ and hence cannot be interior to $\Theta_1$. Therefore,
$l(\Theta_1)=0$, and similarly, $l(\Theta_{k+1})=0$. We thus get
$$\chi(D_{p_1},{\cal O}(D_{p_1}))=\chi(D_{p_{k+1}},{\cal O}(D_{p_{k+1}}))=1,$$
and moreover, $$h^{0,0}=1,\qquad h^{1,0}=0,\qquad h^{2,0}=0,\qquad h^{3,0}=0.$$

For $C_{\tilde p}\cdot C_{\tilde p}=-2$, the points
$p_j, 2 \le j \le k$ lie in a two dimensional
face. Using the methods described above, we find
$$\chi(D_{p_j},{\cal O}(D_{p_j}))=0=-K_{B_2}\cdot C_{\tilde p},$$ and
$$h^{0,0}=1,\qquad h^{1,0}=1,\qquad h^{2,0}=0,\qquad h^{3,0}=0.$$ For the
points $p_1$ and
$p_{k+1}$, we get
$$\chi(D_{p_1},{\cal O}(D_{p_1}))=\chi(D_{p_{k+1}},{\cal O}(D_{p_{k+1}}))=1,$$
and $$h^{0,0}=1,\qquad h^{1,0}=0,\qquad h^{2,0}=0,\qquad h^{3,0}=0.$$

We thus see that the toric method yields results which are in general
agreement with formulas \arthgenB, \arthgenD, and \arthgenF\ with $g(C_{\tilde
p})=0$,
since all the (toric) divisors $C_{\tilde p}$ in the toric variety $B_2$ are
rational.

Another way to compute the arithmetic genus of the horizontal divisors is
to use $\chi(D, {\cal O}(D))= 1/2 K_{B_3}\cdot {\tilde D}^2$ , which
expresses the
arithmetic genus
of the vertical divisor $D$ in the fourfold in terms of its image
${\tilde D}$ in the
base $B_3$, and compute the right hand side from the fan of $B_3$.
As an illustrative example, we explicitly compute the arithmetic genera of
the divisors in the \cyf\ fibered over the base shown in \Blowups (B).
Now, each ray in the fan of the base corresponds to a divisor, and
the canonical class $K=-\Sigma_{i} D_i$. Furthermore, three (distinct)
divisors intersect if and only if they form a cone in the fan. Thus,
for instance, $D_1\cdot D_2\cdot D_5=1$, but $D_5\cdot D_4\cdot D_6=0$. In
addition, there are
three linear relations among the divisors imposed by the structure of the
lattice. These relations are:
\eqn\fanrels{\eqalign{
D_2&=D_4\cr
D_1&=nD_2 + D_7 + D_3 + D_8 + D_9\cr
D_5 + D_7 &= D_8 + 2D_9 + D_6\cr}}

For the fan of \Blowups (A), the corresponding relations are identical except
for the absence of the terms involving $D_7, D_8, D_9$, since these divisors
are absent. In this case, we recognize $D_5,D_6$ as giving the fan of the $P^1$
fiber, and $D_1,\ldots, D_4$ as giving the fan of $F_n$, with $D_2=D_4=f$,
$D_1=C_\infty$, and $D_3=C_0$. (Strictly speaking, $D_1 = P^*(C_\infty)$,
and so on).
The fan of \Blowups (B) then describes the
variety obtained by blowing up $F_n\times P^1$ three times over $C_0$, the
zero section of $F_n$.

After a simple computation, we obtain the following results:
\eqn\kdd{\eqalign{
K\cdot D_1^2&=-2n=-2C_\infty^2,\cr
K\cdot D_2^2&=K\cdot D_4^2=0=-2f^2,\cr
K\cdot D_7^2&=K\cdot D_9^2=2,\cr
K\cdot D_3^2&=K\cdot D_8^2=4-2n=-2K_{F_n}\cdot C_0.\cr}}
The above results again agree with equations
\arthgenA, \arthgenD\ and \arthgenF.

\subsec{Physical Implications}

The above results have interesting physical implications. Note first that 
the behavior of the outer divisors in the chain is different from that 
of the inner divisors. In turn, all inner divisors have the same
number of zero modes. This is actually the expected behavior since 
the outer and inner divisors correspond to two different types of open
membranes as explained above.

Although we have derived a 
necessary and sufficient formula for the generation of a nonperturbative 
superpotential very little is known about its explicit dependence on the 
complex structure moduli. A notable exception is the case treated in 
\ref\DGW{R. Donagi, A. Grassi and E. Witten, ``A Non-Perturbative 
Superpotential With $E_8$ Symmetry'', Mod. Phys. Lett. {\bf A11} (1996) 
2199, hep-th/9607091.}.
Considering an exceptional divisor 
$D_i$ in the chain, the size of the $P^1$ fiber $\phi_i$ combines with 
a Kaluza-Klein mode obtained by reducing the ten dimensional four-form
$C^{(4)}$ on the harmonic form dual to $E_i$ in $B_3$, resulting in a 
chiral multiplet $\Phi_i$. The general expression of the corresponding 
superpotential term is of the form 
\eqn\genexpr{
V\sim e^{-\Phi_i}f(\dots)}
where $f(\ldots)$ is an unknown holomorphic function depending on the 
complex structure moduli of $X$ as well as the $\Phi_j$ and on the 
positions of the background threebranes 
\nref\OG{O. Ganor, ``A Note On Zeroes Of Superpotentials In F-Theory'',
Nucl. Phys. {\bf B499} (1997) 55, hep-th/9612077.}%
\refs{\W,\OG}. 
Furthermore, taking into account the possible extremal transitions 
discussed in the previous section the classical moduli space has a very 
complicated structure. In this case, there could be contributions to the 
superpotential of perturbative nature as explained in the simpler
context of $N=2$ $d=3$ gauge theories in 
\ref\AHISS{O. Aharony, ``Aspects of N=2 Supersymmetric Gauge Theories 
in Three Dimensions'', Nucl. Phys. {\bf B499} (1997) 67, 
hep-th/9703110.}. A discussion of these phenomena in
a particular geometric situation has appeared in 
\ref\DG{D.-E. Diaconescu and S. Gukov, ``Three Dimensional N=2 Gauge 
Theories and Degenerations of Calabi-Yau Four-Folds'', Nucl. Phys. 
{\bf B535} (1998) 17, hep-th/9804059.}. 
Therefore the analysis of this section is a small step towards a complete 
understanding of this complicated moduli space. 

Another interesting aspect is related to the fact that when the number 
of zero modes forbids the generation of a superpotential, there could be 
other associated nonperturbative effects. This is well known in the 
context of $N=1$ field theories with $N_f=N_c$ when such an effect 
results in a quantum deformation of the classical moduli space 
\ref\S{N. Seiberg, ``Exact Results on the Space of Vacua of Four 
Dimensional SUSY Gauge Theories'', Phys.Rev. {\bf D49} (1994) 6857, 
hep-th/9402044.}. As also explained in 
\ref\BSV{M. Bershadsky, `` F-theory, Geometric Engineering and N=1 
Dualities'', Nucl. Phys. {\bf B505} (1997) 153, hep-th/9612052.} 
such a phenomenon can be produced by a 
divisor with $\chi =0$. In the present situation this is obviously the 
case when a horizontal heterotic fivebrane wraps an elliptic curve 
in $C\subset B_2$, according to \arthgenD. This suggests that in this 
case the four dimensional fivebrane--instanton transition can be 
``smoothed'' by nonperturbative effects. However, extra care is needed 
since if $g(C)=1$, we also have $h^{0,1}=1$, hence a cancelation could 
in principle take place.

\bigskip
\noindent {\bf Acknowledgments}
\bigskip
\noindent 
It is a pleasure to thank M. Berkooz, P. Candelas, G. Curio, K. Dasgupta,
A. Grassi, J. Gomis, K. Intriligator, B. Ovrut, N. Seiberg, S. Sethi,
A. Uranga, E. Witten and Z. Yin for useful discussions. The work of
D-E. D. is supported in part by DOE grant DE-FG02-90ER40542. 
The work of G.R.
is supported in part by NSF grant Math/Phys DMS-9627351.

\listrefs
\end